\documentclass[aps,onecolumn,prd,showpacs,showkeys,preprintnumbers,superscriptaddress,nobibnotes,floatfix,longbibliography,notitlepage,nofootinbib]{revtex4-1}

\pdfoutput=1
\usepackage{amsmath}
\usepackage{amsfonts}
\usepackage{amssymb}
\usepackage{mathrsfs}
\usepackage{graphicx}
\usepackage{xcolor}
\usepackage{xspace}
\usepackage{placeins}
\usepackage{fontawesome}

\usepackage{array}
\usepackage{bigints}
\usepackage{tabularx}

\usepackage[colorlinks]{hyperref}
\hypersetup{
    colorlinks,
    linkcolor={red!50!black},
    citecolor={blue!50!black},
    urlcolor={blue!80!black}
}
\usepackage{multirow}
\usepackage{upgreek}
\usepackage[capitalise]{cleveref}
\usepackage{soul}

\newcommand{\eq}{Eq.}
\newcommand{\eqs}{Eqs.}

\newcommand{\Treh}{T_{\mathrm{RH}}\xspace}

\newcommand{\rhobh}{\rho_{\bullet}}

\newcommand{\Mbh}{M_{\bullet}}

\newcommand{\pavg}{\langle p_{dm} \rangle}
\newcommand{\rbh}{r_h}
\newcommand{\Tbh}{T_H}
\newcommand{\nbh}{n_\bullet}
\newcommand{\ndm}{n_{dm}}
\newcommand{\vavg}{\langle v_0 \rangle }
\newcommand{\mWDMmin}{m_{\rm min}^{wdm}}

\begin{document}

\title{Dark Matter from Higher Dimensional Primordial Black Holes}

\author{Avi Friedlander}
\email{avi.friedlander@queensu.ca}
\affiliation{Department of Physics, Engineering Physics and Astronomy, Queen's University, Kingston ON K7L 3N6, Canada}
\affiliation{Arthur B. McDonald Canadian Astroparticle Physics Research Institute, Kingston ON K7L 3N6, Canada}

\author{Ningqiang Song}
\email{songnq@itp.ac.cn}
\affiliation{Institute of Theoretical Physics, Chinese Academy of Sciences, Beijing, 100190, China}
\affiliation{Department of Mathematical Sciences, University of Liverpool, \\ Liverpool, L69 7ZL, United Kingdom}

\author{Aaron C. Vincent}
\email{aaron.vincent@queensu.ca}
\affiliation{Department of Physics, Engineering Physics and Astronomy, Queen's University, Kingston ON K7L 3N6, Canada}
\affiliation{Arthur B. McDonald Canadian Astroparticle Physics Research Institute, Kingston ON K7L 3N6, Canada}
\affiliation{Perimeter Institute for Theoretical Physics, Waterloo ON N2L 2Y5, Canada}


\begin{abstract}
The evaporation of primordial black holes provides a promising dark matter production mechanism without relying on any non-gravitational interactions between the dark sector and the Standard Model. In theories of ``Large'' Extra Dimensions (LEDs), the true scale of quantum gravity, $M_*$, could be well below the Planck scale, thus allowing for energetic particle collisions to produce microscopic black holes in the primordial plasma at temperatures as low as $T \gtrsim 100$ GeV. Additionally, LEDs modify the relationship between black hole mass, radius, and temperature, allowing microscopic black holes to grow to macroscopic sizes in the early Universe. In this work we study three scenarios for the production of dark matter via LED black holes: 1) Delayed Evaporating Black Holes (DEBHs) which grow to macroscopic sizes before ultimately evaporating, 2) Instantly Evaporating Black Holes (IEBHs) which immediately evaporate, and 3) stable black hole relics with a mass $M_*$ known as Planckeons. For a given reheating temperature, $\Treh$, we show that DEBHs produce significantly less dark matter than both IEBHs and Planckeons. IEBHs are able to produce the observed relic abundance of dark matter so long as the reheating scale is in the range $10^{-2} \leq \Treh/M_* \leq 10^{-1}$. We calculate the average speed for the resulting dark matter and show that it would be sufficiently cold for all dark matter masses  $m_{dm} \gtrsim 10^{-4}$ GeV. This mechanism is viable for any scale of quantum gravity in the range $10^4\,\mathrm{ GeV} \leq M_* \leq M_{Pl}$ and for any number of LEDs.
\end{abstract}

\maketitle

\section{Introduction}

It is well established that 85\% of the matter in the Universe is in the form of cold dark matter which primarily interacts gravitationally with the Standard Model \cite{Planck:2018vyg}. The traditional proposal for the nature and origin of dark matter is a Weakly Interacting Massive Particle (WIMP) whose self-annihilations freeze out leaving behind a relic abundance (see Ref.~\cite{Lisanti:2016jxe} for a review). Due to the lack of reproducible direct evidence of WIMP dark matter (see Ref.~\cite{Billard:2021uyg} for a review), it has become increasingly important to study new proposals for both the fundamental nature and production method of dark matter.

As an alternative to thermal production via freeze-out, the Hawking radiation of primordial black holes have often been studied as an intriguing production mechanism for dark matter~\cite{Matsas:1998zm,Bell:1998jk,Khlopov:2004tn,Fujita:2014hha,Allahverdi:2017sks,Lennon:2017tqq,Morrison:2018xla,Hooper:2019gtx,Dalianis:2019asr,Chaudhuri:2020wjo,Hooper:2020evu,Masina:2020xhk,Baldes:2020nuv,Keith:2020jww,Gondolo:2020uqv,Cai:2020kfq,Hooper:2020otu,Bernal:2020kse,Bernal:2020ili,Bernal:2020bjf,Auffinger:2020afu,Kitabayashi:2021hox,Masina:2021zpu,Arbey:2021ysg,Baker:2021btk,Schiavone:2021imu,Sandick:2021gew,Calza:2021czr,Smyth:2021lkn,Barman:2021ost,Samanta:2021mdm,Kitabayashi:2022fqq,Barman:2022gjo,Cheek:2022dbx,Kitabayashi:2022uvu,Mazde:2022sdx,Barman:2022pdo,Bhaumik:2022zdd,Cheek:2022mmy,Agashe:2022phd,Marfatia:2022jiz,Chaudhuri:2023aiv,Cheek:2021cfe,Cheek:2021odj}. Semiclassical arguments suggest that black holes should evaporate to an approximately thermal distribution of particles drawn from  all degrees of freedom that couple to gravity~\cite{Hawking:1975vcx}. This allows primordial black holes to produce dark matter even in the absence of any non-gravitational portals between dark matter and the Standard Model.

While primordial black holes can naturally produce dark matter, the origin of these black holes needs to be explained. The most commonly invoked scenario involves large fluctuations in the primordial power spectrum seeded by inflation, which collapse as they enter the Hubble horizon \cite{Ivanov:1994pa,Garcia-Bellido:1996mdl}. The realization of such curvature fluctuations often involves exotic inflationary scenario such as an ``ultra-slow roll'' phase, whose viability has been recently debated in the context of single-field inflation models~\cite{Riotto:2023hoz,Kristiano:2023scm,Riotto:2023gpm,Franciolini:2023lgy,Firouzjahi:2023ahg} While promising, this mechanism is certainly not the only potential source of primordial black holes.

We have recently demonstrated that primordial black holes can produce the observed dark matter abundance independent of inflationary dynamics during a hot big bang \cite{Friedlander:2023jzw}. This scenario requires that the adiabatic expansion of the Universe began at a temperature within a few orders of magnitude of the Planck scale, hotter than allowed in single-field inflation models \cite{Planck:2018jri}. In this work we extend this proposal to models of ``Large'' Extra Dimensions (LEDs) proposed by Arkani-Hamed, Dimopoulos, and Dvali (ADD) \cite{arkanihamed:1998rs}. This model, originally conceived to address the hierarchy problem, postulates the existence of additional compactified spatial dimensions which would allow the true scale of quantum gravity, $M_*$, to be much lower than the Planck scale.

The properties and hence the cosmological evolution of LED black holes differ substantially from their 4D counterparts~\cite{Argyres:1998qn,Conley:2006jg}. Energetic collisions produce microscopic LED black holes at plasma temperatures close to the reduced quantum gravity scale, which possess a much larger event horizon area than 4D black holes of the same mass, increasing the rate of accretion. The subsequent evolution of LED black holes is subject to two competing effects, accretion and evaporation. Smaller black holes below a threshold mass evaporate immediately; larger ones accrete efficiently and grow in size by many orders of magnitude, surviving over cosmological timescales. Those black holes that survive until today could comprise the dark matter themselves and potentially be detected via astrophysical observations. This was the focus of Refs.~\cite{Johnson:2020tiw,Friedlander:2022ttk}. 

Here, we explore the dark matter production from the Hawking radiation of short-lived Instantly Evaporating Black holes (IEBHs) and longer-lived Delayed Evaporating Black Holes (DEBHs). For both scenarios we find the dark matter mass and the reheating temperature that are compatible with the observed relic abundance of dark matter. IEBHs are able to produce the observed relic abundance of dark matter so long as the reheating scale is in the range $10^{-2} \leq \Treh/M_* \leq 10^{-1}$. Similar to warm dark matter, this relativistically produced dark matter may suppress the structure formation at late times. We calculate the average speed for the resulting dark matter and show that with a mass $m_{dm} \gtrsim 10^{-4}$ GeV, it would be sufficiently cold to evade the warm dark matter constraint. In contrast, DEBHs require a reheating scale of $\Treh/M_*>10^{-1}$ in order to produce the relevant dark matter abundance and will therefore be shown to be subdominant to IEBHs.

In addition, we also explore the possibility of black hole relics being the cosmological dark matter. The semiclassical treatment is expected to break down when the black hole mass approaches the scale of quantum gravity. It is therefore suggested that black holes cease to evaporate and  become stable relics in the last stage of evaporation, known as {\it Planckeons}~\cite{Aharonov:1987tp}. We compute the relic abundance of Planckeons and contrast those with the dark matter produced via black hole evaporation. If Planckeons are stable they can comprise the entirety of dark matter so long as $\Treh=1.5 \times 10^{-2}M_*$ regardless of the number of LEDs.

This article is structured as follows: in Section \ref{sec:LEDPBH} we review the physics of primordial black holes in theories of LEDs including the properties of black holes, their formation, and their evaporation. In Section \ref{sec:cosmology} we describe the cosmological evolution of the black holes and dark matter. Then in Section \ref{sec:results} we present our main results, showing for which reheating temperatures LED black holes are able to produce the observed abundance of dark matter. Lastly we discuss our conclusions and summarize our results in Section \ref{sec:conclusions}.

Except where otherwise stated, we use units where $\hbar=c=k_B = 1$.

\section{Primordial Black Holes in Extra Dimensions} \label{sec:LEDPBH}
\subsection{Black holes in extra dimensions}
In the ADD model, $n$~additional spatial dimensions are compactified with a radius, $R$ \cite{arkanihamed:1998rs}. Over distances much smaller than $R$, the gravitational potential scales more quickly with distance than in 4D Newtonian gravity. The change in scaling implies that the apparent macroscopic strength of gravity, $G = M_{Pl}^{-2}$, is dependent on the true scale of quantum gravity, $M_*$, and the size of the extra dimensions themselves:
\begin{equation}
    M_{Pl}^2\sim M_\star^{2+n}R^n\,.
\end{equation}  
While the precise relationship between $M_*$ and $M_{Pl}$ is dependent on the compactification scheme of the extra dimensions, the effect of the ADD model can be studied by using the Dimopoulos convention~\cite{Argyres:1998qn} where the relationship becomes
\begin{equation}
    M_{Pl}^2\ = M_\star^{2+n}(2\pi R)^n\,.
\end{equation}  
One advantage of this convention is that the standard 4D results for black hole radius, temperature and lifetime can be recovered with $n = 0$ and $M_\star = M_{Pl}$. For $n\geq 2$ LEDs, the strongest direct constraints arise from collider searchers which require that $M_*$ must be at least a few TeV depending on the precise number of LEDs~\cite{Dimopoulos:2001hw,Giddings:2001bu,CMS:2018ucw,CMS:2018ozv}. In the case of $n=1$, the ADD model is most severely constrained by not contradicting macroscopic Newtonian gravity. Torsion balance experiments~\cite{Kapner:2006si} set constraints on the compactification scale of a single LED such that $R\lesssim 1\,\mathrm{ \mu m}$ \cite{Murata:2014nra}. This corresponds to a constraint on the scale of quantum gravity in the case of $n=1$ of $M_* \gtrsim 10^9$ GeV.

The change in gravitational microphysics results in black holes with larger radii $\rbh$ at a given mass than their 4D counterparts when $\rbh \ll R$. In $4+n$ dimensions, black holes have a horizon radius of \cite{Argyres:1998qn}
\begin{equation}
\rbh = \frac{a_n}{M_*} \bigg( \frac{\Mbh}{M_*} \bigg)^{1/(n+1)}\, ,
\end{equation}
where $\Mbh$ is the black hole mass and
\begin{equation}
a_n = \frac{1}{\sqrt{\pi}} \bigg [ \frac{8\Gamma(\frac{n+3}{2})}{n+2}  \bigg]^{1/(n+1)}\,.
\end{equation}
Black holes evaporate via Hawking radiation with a temperature, $\Tbh$, producing all particles that are kinematically allowed, i.e. the particle mass $m\lesssim T_H$ \cite{Hawking:1974rv}. When $r_h\ll R$, the Hawking temperature is related to the horizon radius by \cite{Argyres:1998qn}
\begin{equation}
\Tbh = \frac{n+1}{4\pi \rbh} \,.
\end{equation}

\subsection{Formation of LED black holes}
Energetic particle collisions in the early Universe can produce microscopic black holes when the center-of-mass energy exceeds the quantum gravity scale, turning the center-of-mass energy into the black hole mass $\Mbh$. During radiation domination, the production rate for microscopic black holes can be computed by convolving the black hole production cross section with the particle distribution functions in the plasma, resulting in~\cite{Friedlander:2022ttk}
\begin{equation}
    \dfrac{d\Gamma}{d\Mbh}=\dfrac{g_{\star}(T)^2a_n^2}{8\pi^3}\Mbh T^2\left(\dfrac{\Mbh}{M_\star}\right)^{\frac{2n+4}{n+1}}\left[\dfrac{\Mbh}{T}K_1(\frac{\Mbh}{T})+2K_2(\frac{\Mbh}{T})\right]\Theta(\Mbh-M_\star)\,,
    \label{eq:BHprodrate}
\end{equation}
where $g_*(T)$ is the number of effective degrees of freedom in the plasma with a temperature $T$, $K_i(x)$ is the modified Bessel function of the second kind, and $\Theta(x)$ is the Heaviside step function. After formation, the black holes undergo two competing processes: accretion of the surrounding plasma and evaporation to Hawking radiation. Accreting black holes grow at a rate of
\begin{equation}
    \dfrac{d\Mbh^\mathrm{acc}}{dt}=\beta\dfrac{T^4}{\Tbh^2}\,,
    \label{eq:dMdtcombine}
\end{equation}
where
\begin{equation}
\beta = \frac{\pi(n+1)^2 f_\mathrm{acc} g_*(T)}{120}\,,
\end{equation}
and $f_{\rm acc}$ is an $\mathcal{O}(1)$ constant that describes the accretion efficiency. In this work we assume $f_{\rm acc}=1$. Simultaneously, the $M_*$-scale black holes evaporate at a rate of 
\begin{equation} \label{eq:dMdTHighTH}
\frac{d M_{\bullet}^\mathrm{evap}}{dt} = - \alpha_0 \Tbh^2\,,
\end{equation}
where $\alpha_0$ is a constant that only depends on $n$ and will be further defined in Section~\ref{sec:Evap}.

Primordial black holes formed via particle collisions can be split into two categories based on whether evaporation or accretion dominates the initial black hole evolution. {\it Instantly Evaporating Black Holes} (IEBHs) are produced with masses below a threshold mass, $M_{th}$, such that they evaporate faster than they accrete. Alternatively, {\it Delayed Evaporating Black Holes} (DEBHs) form with a mass $\Mbh > M_{th}$, such that they grow before finally evaporating. The threshold mass above which black holes grow from their initial size is
\cite{Friedlander:2022ttk}
\begin{equation}
    M_{th}=\max\left\{M_\star,\left[\dfrac{n+1}{4\pi a_n}\left(\dfrac{\alpha_0}{\beta}\right)^{1/4}\dfrac{M_\star}{T}\right]^{n+1}M_\star\right\}\,.
    \label{eq:Mth}
\end{equation}

The total production rate of black holes per unit volume can be expressed as 
\begin{equation}
    \Gamma_\bullet = \Gamma_I + \Gamma_D\,,
\end{equation}
where $\Gamma_I$ and $\Gamma_D$ are the production rates of IEBHs and DEBHs respectively. While the production of both types of black holes could occur simultaneously, we will show that the relevant temperatures at which IEBHs and DEBHs would produce the observed abundance of dark matter does not overlap. Therefore, we will study them independently so that their impact can be easily compared.

The production rate for IEBHs is
\begin{equation} \label{eq:gammaIEBH}
\Gamma_I = \int_{M_*}^{M_{th}} d\Mbh \dfrac{d\Gamma}{d\Mbh} \,,
\end{equation}
and for DEBHs
\begin{equation} \label{eq:gammaDEBH}
\Gamma_D = \int_{M_{th}}^\infty d\Mbh \dfrac{d\Gamma}{d\Mbh} \,.
\end{equation}
In the limit of instantaneous black hole formation, the initial number density of black holes is
\begin{equation}
\nbh(\Treh) \approx \int_{\Treh}^0 dT  \left(\frac{dT}{dt}\right)^{-1} \; \Gamma_D(T) \,,
\label{eq:nbhReh}
\end{equation}
where $\Treh$ is the initial temperature at which the adiabatic expansion of the Universe begins, typically referred to as the {\it reheating temperature}. Only DEBH production contributes to the density of black holes because the IEBHs evaporate before any significant density accumulates.

\subsection{Evaporation of LED black holes} \label{sec:Evap}
Black holes evaporate to all particle degrees of freedom based on the horizon radius of the black hole and the phase-space available to the evaporation product. The rate of mass loss from evaporation to species $i$ can be compactly expressed as
\begin{equation}
    \frac{d M_{\bullet\to i}}{dt} = \dfrac{-g_{i}}{2\pi}\dfrac{\xi_i}{r_h^2}\,,
    \label{eq:dMdtSpecies}
\end{equation}
where $g_{i}$ is the number of degrees of freedom for species $i$. If the particles are emitted on the 4D brane, $\xi_i$ is the dimensionless integral of the Bose-Einstein or Fermi-Dirac distribution weighted by particle energy and a greybody distortion factor $\sigma_i$, i.e.
\begin{equation} \label{eq:xiDef}
    \xi_i=\int \dfrac{\sigma_i}{\pi r_h^2}\dfrac{E r_h^4}{\exp(E/T_H)\mp 1}p^2dp\,,
\end{equation}
where $E$ and $p$ are respectively the energy and momentum of the evaporation product. While most Standard Model particles are confined to the 4D brane, gravitons are able to travel throughout the (4$+n$)D bulk thus enlarging the phase space of the graviton evaporation products. Evaporation to gravitons is instead formulated by the momentum integral
\begin{equation} \label{eq:xiGrav}
    \xi_G=\int \sum_l N_l |A_l|^2 \dfrac{E r_h^2}{\exp(E/T_H)- 1}dp\,,
\end{equation}
where $A_l$ is the absorption coefficient which describes the ratio of radiation in quantum state $l$ at infinity to that at the horizon and $N_l$ is the multiplicity of state $l$. The integral in \eq~\eqref{eq:xiGrav} accounts for all degrees of freedom of gravitons so for the purposes of computing the evaporation rate in \eq~\eqref{eq:dMdtSpecies}, we treat $g_\mathrm{graviton}=1$. Semi-analytic fits for $\xi_i$ and $\sigma_i$ for LED black holes are provided in Ref. \cite{Friedlander:2022ttk} along with numerical results for $\sum N_l|A_l|^2$ in the case of gravitons. The total evaporation rate for a black hole is found by summing \eq~\eqref{eq:dMdtSpecies} over all particle species:
\begin{equation}
    \frac{d M_{\bullet}^\mathrm{evap}}{dt} = \sum_i \dfrac{d M_{\bullet\to i}}{dt}\,.
    \label{eq:dMdtEvap}
\end{equation}
In the limit where the black holes are light enough such that $T_H \gg m_i$ for all emitted particles, the evaporation rate reduces to \eq~\eqref{eq:dMdTHighTH} with
\begin{equation}
    \alpha_0 \equiv \frac{8\pi}{(n+1)^2}( g_0 \tilde{\xi}_0 + g_{1/2} \tilde{\xi}_{1/2} + g_1 \tilde{\xi}_{1} + \xi_G) \,,
\end{equation}
where $g_0$, $g_{1/2}$ and $g_1$ are the number of degrees of freedom of scalars, spinors, and vectors respectively and $\tilde{\xi}_i$ is the massless limit ($E = p$) of the phase-space integral \eqref{eq:xiDef} for a species with spin $i$. Numerical values for $\tilde{\xi}_i$ and $\alpha_0$ for all $n$ are provided in Table \ref{tab:xiEtaValues}

\begin{table}[htb]
\begin{center} 
\setlength\extrarowheight{3pt}
\begin{tabular}{c | c c  c  c | c c c | c } 
 \hline \hline
  n & $\tilde{\xi}_0$ & $\tilde{\xi}_{1/2}$ & $\tilde{\xi}_1$ &  $\tilde{\xi}_G$ & $\tilde{\eta}_0$ & $\tilde{\eta}_{1/2}$ & $\tilde{\eta}_1$ & $\alpha_0$\\ [0.5ex] 
 \hline
 0 & $1.87\times10^{-3}$  &  $1.03\times10^{-3}$ & $4.23\times10^{-4}$ & $9.66\times10^{-5}$ &  $8.36\times10^{-3}$ &  $3.05\times10^{-3}$ &  $9.30\times10^{-4}$ & $2.77$\\ 
 1 & $1.67\times10^{-2}$  & $1.46\times10^{-2}$  & $1.15\times10^{-2}$ & $9.71\times10^{-3}$ & $3.97\times10^{-2}$ & $2.76\times10^{-2}$ & $1.78\times10^{-2}$ & 10.45\\ 
 2 & $6.75\times10^{-2}$  &  $6.12\times10^{-2}$ & $6.11\times10^{-2}$ & $9.95\times10^{-2}$ & 0.109 &  $8.41\times10^{-2}$ &  $7.49\times10^{-2}$ & $20.50$\\ 
 3 & 0.187 &  0.167 & 0.186 & 0.493 & 0.229 & 0.178 & 0.189 & $32.53$\\ 
 4 & 0.416  & 0.362  & 0.432 & 1.90 & 0.412 & 0.313 & 0.374 & $46.74$\\ 
 5 & 0.802  &  0.684 & 0.847 & 6.89 & 0.668 & 0.496 & 0.640 & $64.19$\\ 
 6 & 1.40  &  1.17 & 1.49 & 24.7 & 1.01 & 0.731 & 0.997 & $88.03$\\ 
 \hline \hline
\end{tabular}
\end{center}
\caption{Numerical values for $\xi_i$ and $\eta_i$ for each particle spin and number of LEDs in the limit where the mass of the emitted particle $m\ll \Tbh$. The last column, $\alpha_0$, accounts for all Standard Model degrees of freedom and gravitons.}
\label{tab:xiEtaValues}
\end{table}

\begin{figure}[htb]
    \centering
    \includegraphics[width=0.495\textwidth]{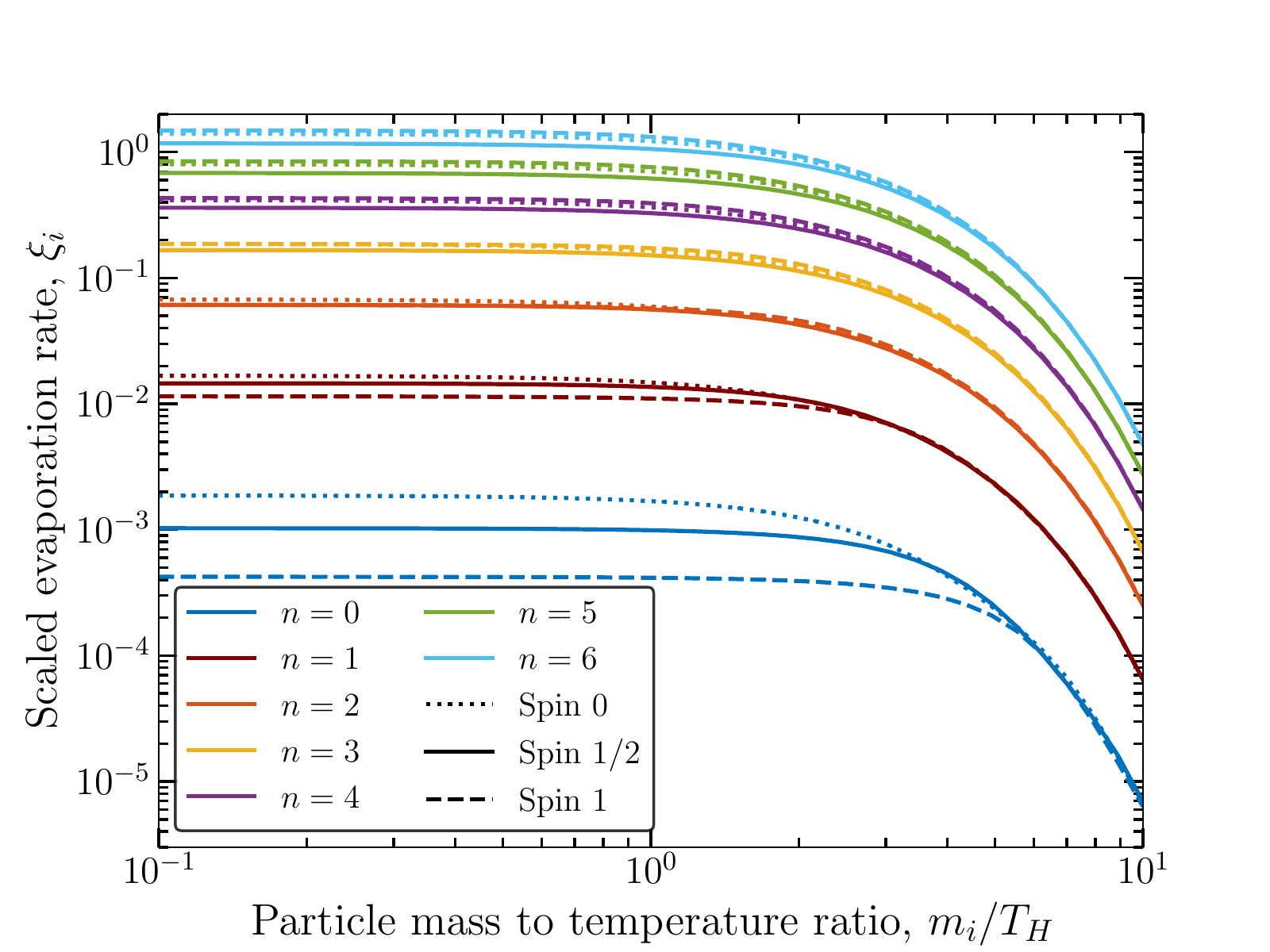}
    \includegraphics[width=0.495\textwidth]{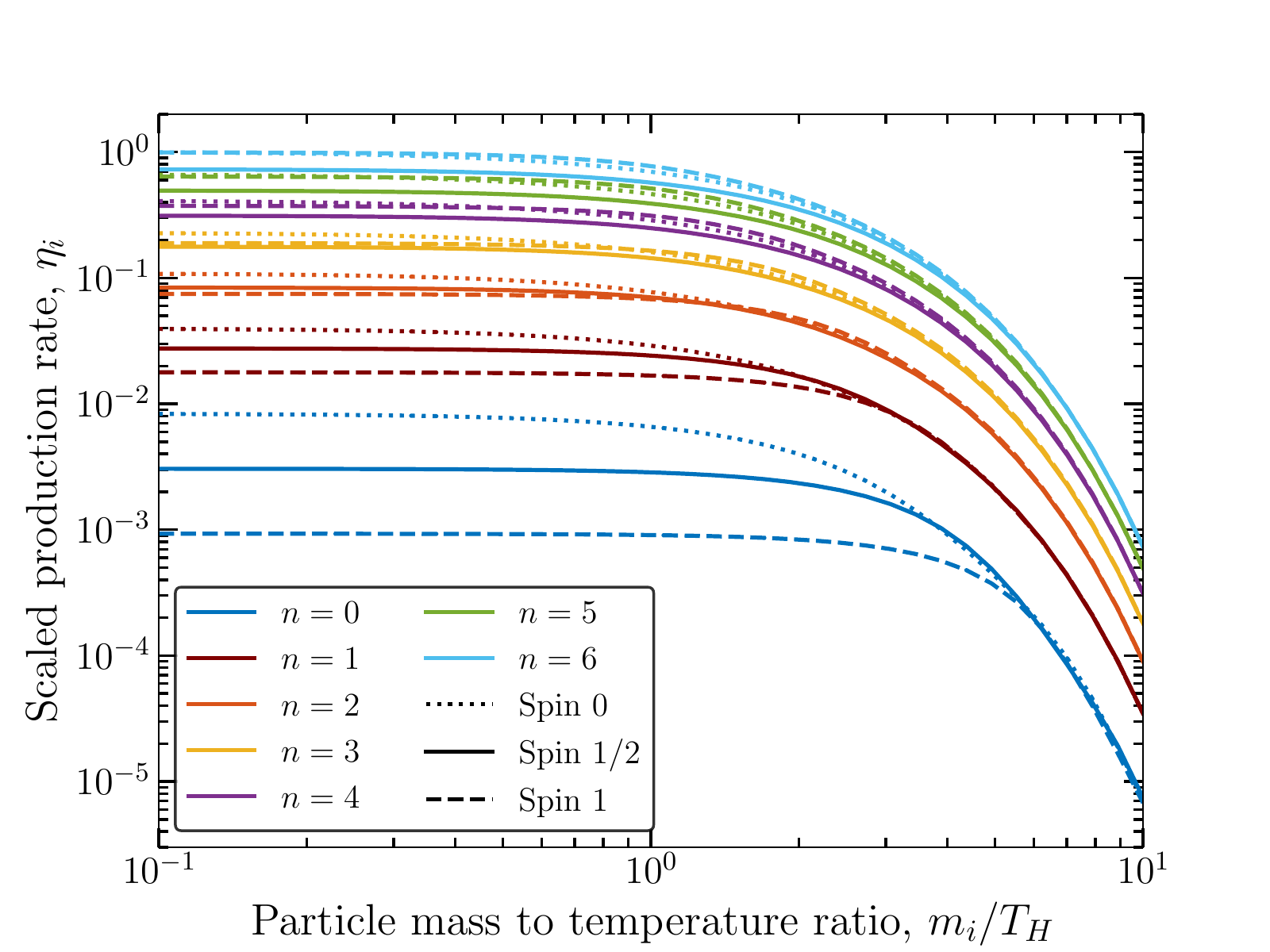}
    \caption{Dependence of the evaporation and particle production rates on the ratio of the mass of the evaporation product, $m_i$, to the Hawking temperature of the black hole, $\Tbh$. The different line colours represent different numbers of LEDs and the dotted, solid, and dashed curves represent evaporation to a scalar, spinor, and vector respectively. In the limit of $m_i/\Tbh \ll 1$, the values of $\xi_i$ and $\eta_i$ approach the results presented in Table~\ref{tab:xiEtaValues}. Left: the scaled evaporation rate to a given species as defined in \eqs~\eqref{eq:dMdtSpecies} and \eqref{eq:xiDef}. Right: the scaled rate of particle production as defined in \eqs~\eqref{eq:partProdRate} and \eqref{eq:etaDef}.}
    \label{fig:xietaVals}
\end{figure}

The particle production rate from an evaporating black hole can be found similarly to the mass-loss rate, except without the energy weighting in the integrand of \eq~\eqref{eq:xiDef}. Therefore, the number of particles of species $i$ produced by a single black hole per unit time is
\begin{equation} \label{eq:partProdRate}
\frac{dN_i}{dt} = \frac{g_{i}}{2\pi}\frac{\eta_i}{ \rbh} \,,
\end{equation}
where
\begin{equation} \label{eq:etaDef}
\eta_i=\int \dfrac{\sigma_i}{\pi r_h^2}\dfrac{r_h^3}{\exp(E/T_H)\mp 1}p^2dp\,.
\end{equation}
Figure \ref{fig:xietaVals} shows both $\xi_i$ and $\eta_i$ evaluated as a function of the mass of the evaporation product, $m_i$. When $\Tbh \gg m_i$, the phase-space integral in \eq~\eqref{eq:etaDef} approaches a constant, $\tilde{\eta}_i$. Numerical values for $\tilde{\eta}_i$ for different spin and numbers of extra dimensions are shown in Table \ref{tab:xiEtaValues}.

\section{Cosmological Evolution} \label{sec:cosmology}
After black holes form, their cosmological evolution differs significantly depending on whether they are DEBHs, growing to larger masses and surviving over cosmological timescales, or IEBHs, which decay and disappear instantly. We use different methodologies to analyze the impact of each scenario. In the case of DEBHs we numerically solve coupled differential equations tracking the growth and evaporation of black holes simultaneously, while in the case of IEBHs we are able to make simplifying assumptions leading to analytic expressions for the relic abundance of dark matter. In this section we present the methodology for each scenario.

\subsection{Delayed Evaporating Black Holes}

The evolution of DEBHs in the early Universe can be connected with observations of the late Universe by establishing a consistent system of differential equations that tracks the black hole, radiation, and dark matter energy densities over time. In doing so we closely follow the methodology of previous work which has studied black hole evaporation to dark matter~\cite{Cheek:2021odj}, but with the modified production, accretion, and evaporation due to the LEDs. We will first justify treating the DEBHs as having a monochromatic mass spectrum by deriving an analytic approximation of the maximum mass that DEBHs grow to before describing the full evolution of DEBHs.

We treat all DEBHs as having the same mass. This is a good approximation for $n\geq2$ because all DEBHs will grow to a similar maximum mass, $M_\mathrm{max}$, regardless of their size at formation and with only weak dependence on the plasma temperature at formation, $T_i$~\cite{Friedlander:2022ttk}. An approximation for $M_\mathrm{max}$ can be found by ignoring the black hole's evaporation and assuming only accretion is important for the first stages of their evolution. If $n\geq2$ and DEBHs grow only in a radiation dominated phase, then all DEBHs will have a mass close to the asymptotic mass
\begin{equation}
   M_{as}= \left(\gamma_n\dfrac{M_{Pl}T_i^2}{M_\star^3}\right)^\frac{n+1}{n-1}M_\star\,,
   \label{eq:asympmass}
\end{equation}
where
\begin{equation}
\gamma_n=f_{\rm acc}a_n^2\sqrt{\frac{\pi^3g_\star}{20}}\frac{n-1}{n+1} \,.
\end{equation}
If DEBHs accrete an $\mathcal{O}(1)$ fraction of the radiation density such that they come to dominate the Universe, the radiation temperature will rapidly drop and accretion will shut off. This results in a reduction of $M_\mathrm{max}$ relative to $M_{as}$:
\begin{equation} \label{eq:Mmax}
M_{\mathrm{max}}= \bigg\{
   \begin{array}{lr}
       M_{as}, & M_{as}\nbh(\Treh) \le \rho_r(\Treh) \,,\\
        \rho_r(\Treh)/\nbh(\Treh) & M_{as}\nbh(\Treh) > \rho_r(\Treh) \,,
   \end{array}
\end{equation}
where $\rho_r$ is the radiation energy density. This reflects a somewhat different approach than we previously took in Ref.~\cite{Friedlander:2022ttk}. In Ref.~\cite{Friedlander:2022ttk} for cases where black holes would accrete the majority of the radiation density, we assumed that the number of formed black holes is reduced so that the maximum mass they grow to is always $M_{as}$. Both approaches only approximate the full solution which includes simultaneous black hole formation and accretion. However, so long as $n_\bullet(\Treh)M_{th} \ll \rho_r(\Treh)$, the timescale of black hole formation is shorter than the timescale of accretion and the number density of formed DEBHs would not be cutoff.

The fact that all DEBHs can be treated with a single mass can be understood from the fact that \eq~\eqref{eq:asympmass} is independent of  $M_i$, the mass of the black hole before accretion, and only has a polynomial dependence on the formation temperature. This is not the case for $n=1$ where $M_{as}$ can be found to be
\begin{equation} \label{eq:n1Mas}
M_{as} = M_i e^{\gamma_1\frac{M_{Pl}T_i^2}{M_*^3}} \,,
\end{equation}
where
\begin{equation}
    \gamma_1= f_\mathrm{acc}a_1^2 \sqrt{\frac{\pi^3 g_*}{20}} \,.
\end{equation}
Since $M_{as}$ is proportional to $M_i$ and depends exponentially on the ratio $M_{Pl}T_i^2/M_*^3$,
DEBHs with $n=1$ will either all grow large enough to saturate the bulk and act as long-lived macroscopic 4D black holes or they will have an extended mass spectrum. In either case, the methodology presented in this  section would not apply. We therefore focus on $n\geq2$ and leave an analysis of DEBHs with $n=1$ to future work. 

\begin{figure}[htb]
    \centering
    \includegraphics[width=0.495\textwidth]{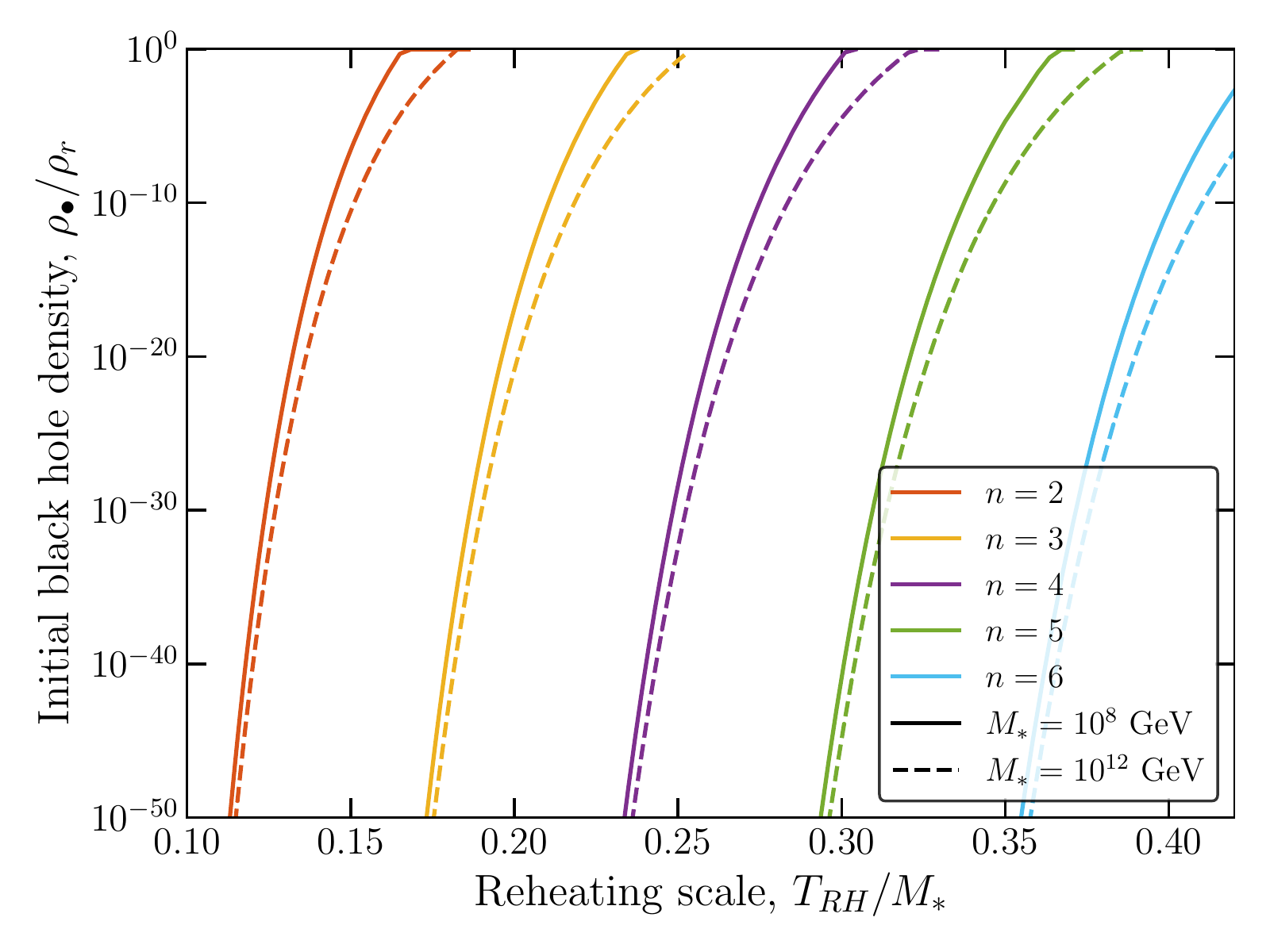}
    \includegraphics[width=0.495\textwidth]{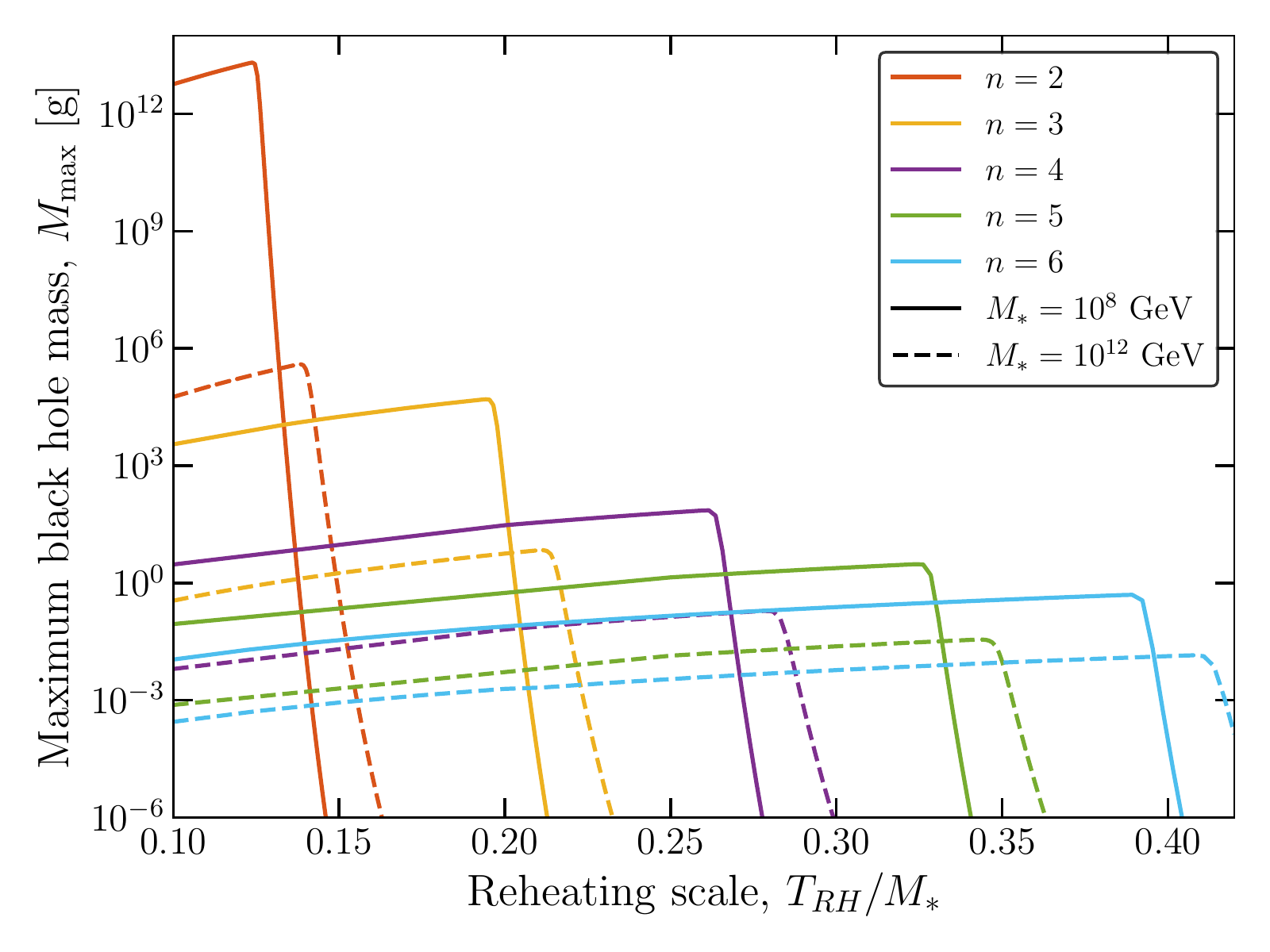}
    \caption{Properties of DEBHs as a function of the reheating scale, $\Treh/M_*$. Different colour curves depict different numbers of LEDs. The solid curves have a scale of quantum gravity of $M_* = 10^8$ GeV whereas the dashed curves show  $M_* = 10^{12}$ GeV. Left: The approximate fraction of the Universe which initially forms DEBHs assuming all black holes form with $\Mbh=M_{th}$. Right: The maximum mass DEBHs grow to before fully evaporating.}
    \label{fig:productionResults}
\end{figure}

Figure \ref{fig:productionResults} shows how the number density of the produced DEBHs and the maximum mass they grow to depends on the fundamental LED model parameters. The left panel shows the fraction of the radiation density which initially converts to DEBHs via particle collisions assuming that all DEBHs form with a mass of $\Mbh = M_{th}$. With this assumption the initial density of black holes is given by $\rhobh = \nbh M_{th}$. When $\rhobh \sim \rho_r$, black hole formation will deplete the available plasma resulting in a more complicated temperature evolution when calculating $\nbh$ with \eq~\eqref{eq:nbhReh}. However, for scenarios where the observed relic abundance of dark matter is produced, this is not an issue because $\rhobh \ll \rho_r$. The right panel of Figure \ref{fig:productionResults} shows the maximum mass that DEBHs grow to as a function of the reheating scale.  For smaller values of $\Treh$ the maximum DEBH mass mostly depends on the features of the LED model: $n$ and $M_*$. However, as more DEBHs form at larger $\Treh$, the maximum mass DEBHs reach sharply decreases because they accrete the majority of the radiation in the Universe creating an early matter dominated era before evaporating back into radiation. While this figure was produced by numerically solving the full system of equations described below, this behaviour exactly matches the prediction for $M_\mathrm{max}$ described in \eq~\eqref{eq:Mmax}.

To produce analytic approximations for properties such as $M_\mathrm{max}$ it can be helpful to separate DEBH evolution into two distinct phases where first accretion occurs followed by evaporation. However, both processes do occur simultaneously. We account for this by numerically solving a system of differential equations which consistently accounts for the complete evolution of DEBHs and their interaction with the plasma. Therefore, we use the full black hole mass evolution which can be expressed as:
\begin{equation} \label{eq:dMdtfull}
\frac{d\Mbh}{dt} = \frac{d M_{\bullet}^\mathrm{evap}}{dt}  + \dfrac{d\Mbh^\mathrm{acc}}{dt}\,.
\end{equation}

The number density of black holes, $\nbh$ is diluted by the homogeneous expansion of the Universe:
\begin{equation} \label{eq:dnbhdt}
\frac{d\nbh}{dt} = -3H\nbh \,,
\end{equation}
where the Hubble rate is
\begin{equation} \label{eq:Hubble}
H = \frac{1}{M_{Pl}} \sqrt{\frac{8\pi}{3}(\rho_r + \rho_\bullet + \rho_{dm} + \rho_\Lambda)} \,,
\end{equation}
and $\rho_i$ is the energy density of each component of the Universe. In the case of black holes, this is simply $\rho_\bullet = \Mbh \nbh$. For radiation, the energy density is
\begin{equation}
\rho_r = \frac{\pi^2}{30}g_* T^4 \,,
\end{equation}
where $T$ is the plasma's temperature. The plasma temperature evolves as
\begin{equation} \label{eq:dTdt}
\frac{dT}{dt} = -\frac{T}{1 + \frac{T}{3} \frac{d\ln(g_{*s})}{dT} } \bigg[H + \frac{\nbh}{4\rho_r}\bigg(\frac{d\Mbh}{dt} - \frac{d M_{\bullet\to dm}}{dt}\bigg)\bigg]\,,
\end{equation}
where $g_{*s}$ is the number of effective degrees of freedom which contribute to entropy. The first term in brackets accounts for Hubble expansion while the second accounts for interactions between DEBHs and the radiation bath. Black hole accretion reduces the radiation density while evaporation deposits energy back into the plasma. Therefore, the change in temperature is dependent on the change in the DEBHs' mass excluding mass change caused by evaporation to dark matter. Lastly, when relativistic Standard Model species freeze out, they deposit their entropy into the radiation sector. This effect is included in the denominator of the prefactor. 

To study the evolution of dark matter, assumptions must be made about its fundamental nature. We focus on a minimalistic dark matter model where all of the dark matter is comprised of a single species that is stable over cosmological time scales and only interacts with the Standard Model via gravity. In this case, the number density of dark matter, $\ndm$, is only affected by the expansion of the Universe and black hole evaporation such that
\begin{equation} \label{eq:dndmdt}
\frac{d\ndm}{dt} = -3H\ndm + \nbh \frac{dN_{dm}}{dt} \,.
\end{equation}

 Since these dark matter particles are produced relativistically, they may suppress the structure formation in a similar manner as warm dark matter. Therefore, it is important to examine whether the produced dark matter is cold enough to evade the constraints on small scale structures. To compare the dark matter produced from DEBHs with limits on its velocity it is important to have information about the evolution of the momentum distribution of dark matter. We do not track the full phase-space distribution but rather determine the average momentum of dark matter particles, $\pavg$. Similar to the evolution of $n_{dm}$, in the absence of inelastic interactions or interactions with the Standard Model, the momentum energy density of dark matter, $\pavg \ndm$, is only impacted by the expansion of the Universe and black hole evaporation:
\begin{equation} \label{eq:pavgEvolution}
\frac{d(\pavg \ndm)}{dt} = -4 H \pavg \ndm + \frac{dM_{\bullet\to dm}}{dt} n_\bullet \,.
\end{equation}
A derivation of this relationship can be found in Appendix \ref{app:momEvolDeriv}. The total dark matter energy density in \eq~\eqref{eq:Hubble} is therefore tracked as $\rho_{dm} \approx \ndm\sqrt{\pavg^2 + m_{dm}^2}$. However, dark matter which has the correct cosmological abundance and is sufficiently cold today, will have a negligible average momentum during any epoch where $\rho_{dm}$ comprises an $\mathcal{O}(1)$ fraction of the Universe. Therefore, the kinetic energy contribution to the energy density can be safely ignored such that $\rho_{dm} \approx n_{dm} m_{dm}$.

\eqs~\eqref{eq:dMdtfull}, \eqref{eq:dnbhdt}, \eqref{eq:dTdt}, \eqref{eq:dndmdt}, and \eqref{eq:pavgEvolution} form a system of differential equations governing the evolution of DEBHs, radiation, and dark matter in an expanding Universe. To connect a theory of extra dimensions to a cosmological history, this system of equations can be solved with initial conditions such that at a reheating temperature, $\Treh$, the Universe solely consists of radiation and a population of black holes whose mass is the threshold mass \eqref{eq:Mth} and have a number density described by \eq~\eqref{eq:nbhReh}. 

\begin{figure}[htb]
    \centering
    \includegraphics[]{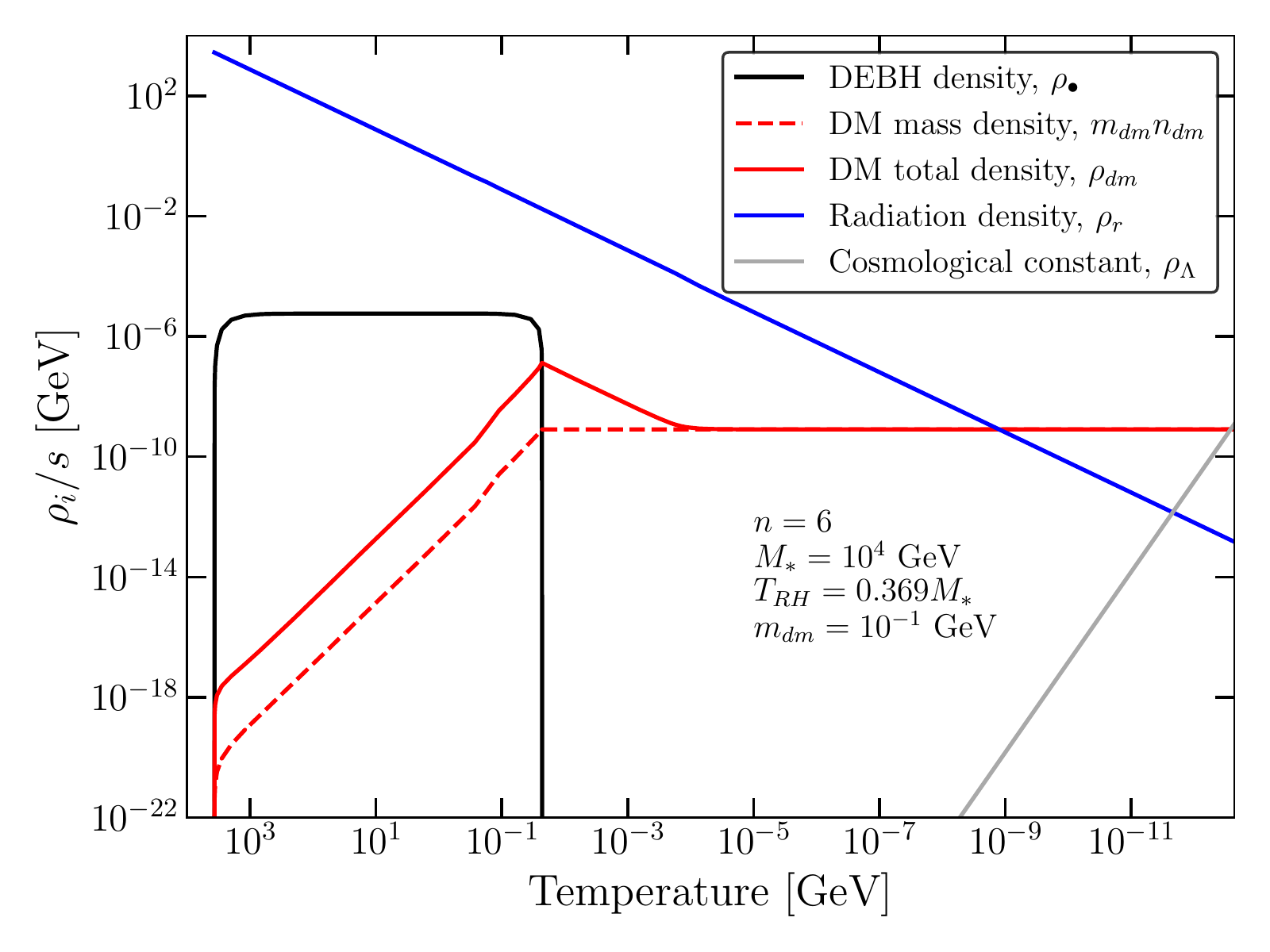}
    \caption{The cosmological history of DEBHs in a Universe with 6 additional spatial dimensions, $M_* = 10^4$ GeV, and $\Treh = 0.369 M_*$ where the entirety of dark matter today is comprised of a $10^{-1}$ GeV mass Dirac fermion. Each line depicts the evolution of a cosmological component's density divided by the radiation entropy. The blue line depicts the radiation density, the grey line depicts the cosmological constant, the black curve depicts the black hole energy density, and the red curve depicts the dark matter (DM) energy density. The red dashed curve shows the dark matter energy density ignoring any kinetic energy the particles have at a given time.}
    \label{fig:History}
\end{figure}

Figure \ref{fig:History} shows an example of a cosmological history with DEBHs which evaporate to a $0.1$ GeV mass Dirac fermion dark matter particle (red). In this example, the black holes (black) completely evaporate early in the Universe without ever dominating over the radiation density. While the dark matter is produced relativistically, it cools well before matter-radiation (blue) equality; this can be seen in the difference between the total dark matter energy density (solid red) and the energy density in mass alone (dashed red). 

\subsection{Instantly Evaporating Black Holes} \label{sec:IEBHanyl}
In the case of IEBHs, the resulting energy density of dark matter can be obtained analytically. In this section we follow the approach of Ref.~\cite{Friedlander:2023jzw} in which we obtained analytic expressions for IEBHs in a 4D Universe, and generalize the derivation to a Universe with extra spatial dimensions.

The total number density of IEBHs is well approximated by assuming they form at a temperature $T \ll M_\star$ and during radiation domination. In this regime, the differential production rate of black holes from \eq~\eqref{eq:BHprodrate} can be approximated by using the fact that $K_i(x) \approx \sqrt{\pi/(2x)}e^{-x}$ for large $x$ resulting in
\begin{equation}
 \dfrac{d\Gamma}{d\Mbh}=\dfrac{g_{\star}(T)^2a_n^2}{8\pi^3}\sqrt{\frac{\pi }{2}}(\Mbh T)^{3/2}\left(\dfrac{\Mbh}{M_\star}\right)^{\frac{2n+4}{n+1}}e^{-\Mbh/T} \Theta(\Mbh-M_\star)\,.
 \label{eq:bhProdRateLowT}
\end{equation}

Most black holes have mass $\Mbh\sim M_*$ at formation, corresponding to a Hawking temperature that is above the mass of all fundamental particles. The total number of dark matter particles produced per black hole formed, $N_{dm}$, can be obtained analytically as
\begin{align}
    N_{dm}(\Mbh) &= \int_{\Mbh}^0 d\Mbh^\prime \bigg(\frac{d\Mbh^\prime}{dt}\bigg)^{-1}\frac{dN_{dm}}{dt} \nonumber\\
    &= \frac{8\pi a_n g_{dm} \tilde{\eta}_i}{(1+n)(2+n)\alpha_0}\left( \frac{\Mbh}{M_*}\right)^{\frac{2+n}{1+n}}\,,
\end{align}
where $d\Mbh/dt$ only accounts for evaporation and is approximated using the high-$\Tbh$ limit described by \eq~\eqref{eq:dMdTHighTH} and $dN_{dm}/dt$ is found using the particle production rate \eqref{eq:partProdRate} in the massless product limit.

The number density of dark matter at production is found by convolving the production of dark matter from black holes with the production of black holes over all temperatures
\begin{align}
n_{dm}(\Treh) &= \int_{\Treh}^0 dT \bigg(\frac{dT}{dt}\bigg)^{-1} \int_{M_*}^\infty d\Mbh N_{dm}(\Mbh) \frac{d\Gamma}{d\Mbh} \label{eq:ndmIEBHIntegral}\\
 &= \frac{3\sqrt{\frac{5}{2}} a_n^3 g_*^{3/2} g_{dm} \tilde{\eta}_i M_{Pl} M_*^2}{2(2+n)(8+5n)\pi^3 \alpha_0} \left[\left(\frac{\Treh}{M_*}\right)^\frac{8+5n}{1+n} \Gamma\left(\frac{17 + 11n}{2+2n}, \frac{M_*}{\Treh} \right) - \Gamma\left(\frac{1}{2}, \frac{M_*}{\Treh} \right)\right]\,, \label{eq:ndmIEBHAnyl}
\end{align}
where the temperature evolution does not account for any accretion so $dT/dt = -HT$ and $\Gamma(s,x)$ is the upper incomplete gamma function. While \eq~\eqref{eq:ndmIEBHAnyl} is the analytic expression used to obtain the results throughout this work, it can be approximated to leading order in $\Treh/M_\star$ as
\begin{equation}
    n_{dm}(\Treh) \approx \frac{3\sqrt{\frac{5}{2}} (16+10n) a_n^3 g_*^{3/2} g_{dm} \tilde{\eta}_i M_{Pl} M_*^2}{4(1+n)(2+n)(8+5n)\pi^3 \alpha_0} \bigg(\frac{\Treh}{M_*}\bigg)^{3/2} e^{M_*/\Treh}\,,
\end{equation}
indicating that the number density of dark matter depends exponentially on $M_\star/\Treh$.

Assuming that after production the dark matter does not significantly interact with the Standard Model contents of the Universe, the ratio of dark matter number density to entropy remains constant. Therefore the relic abundance today is 
\begin{equation} \label{eq:OmDMIEBH}
    \Omega_{dm} = \frac{g_{*s}(T_0) T_0^3 m_{dm} n_{dm}(\Treh) }{g_{*s}(\Treh) \Treh^3 \rho_{c}}\,,
\end{equation}
where $\rho_c$ is the critical density of the Unvierse today and $T_0$ is the CMB temperature today. For all the results presented in this work, we have used the full analytic expression for $n_{dm}(\Treh)$ shown in \eq~\eqref{eq:ndmIEBHAnyl} rather than the leading order approximation.

\begin{figure}[htb]
    \centering
    \includegraphics[width=0.49\textwidth]{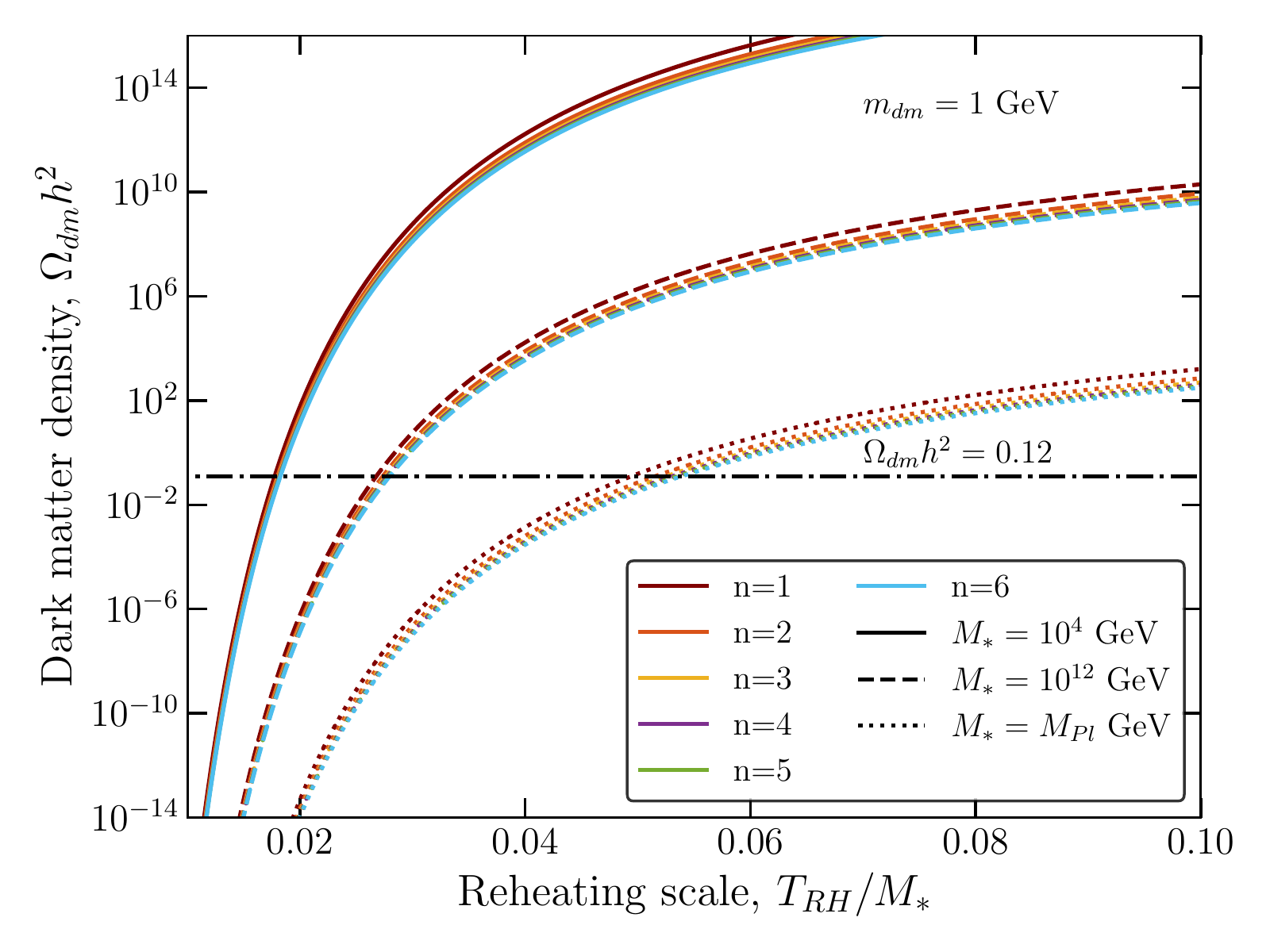}
    \includegraphics[width=0.49\textwidth]{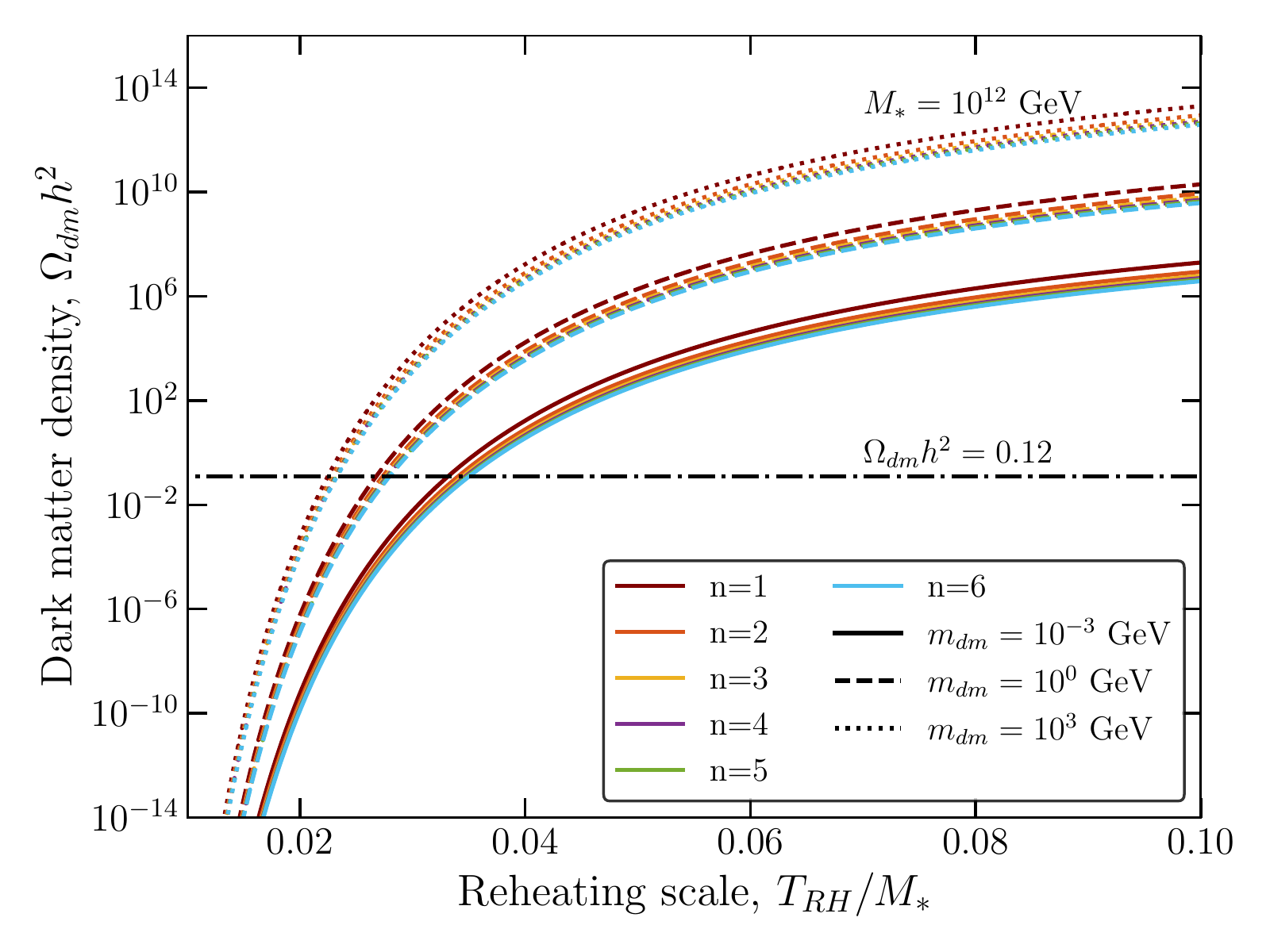}
    \caption{Relic abundance of Dirac fermion dark matter produced by IEBHs. Curves of different colors represent different number of extra dimensions. The black dot-dashed line shows the observed relic density of dark matter today. Left: Dark matter mass is fixed at $m_{dm} = 1$ GeV for all curves. The solid, dashed, and dotted curves show scales of quantum gravity of $10^4$ GeV, $10^{12}$ GeV, and $M_* = M_{Pl}$ respectively. Right: The scale of quantum gravity is fixed at $M_* = 10^{12}$ GeV for all curves. The solid, dashed, and dotted curves show the relic abundance for dark matter with masses of $10^{-3}$ GeV, $1$ GeV, and $10^3$ GeV respectively.}
    \label{fig:IEBHdensity}
\end{figure}

Figure \ref{fig:IEBHdensity} shows how the relic density of Dirac fermion dark matter produced by IEBHs depends on $\Treh/M_*$ assuming that the initial plasma consists of only Standard Model particles. The different colour curves represent different numbers of extra dimensions demonstrating that for a fixed $M_*$, $\Treh$, and dark matter model the number of extra dimensions has a minimal impact on the resulting relic density. The left panel fixes the dark matter mass at $m_{dm} = 1$ GeV while varying $M_*$, showing that when $M_*$ increases, $\Treh$ is required to be closer to $M_*$ to achieve the same relic density. Alternatively, the right panel fixes the scale of quantum gravity at $M_* = 10^{12}$ GeV and shows that heavier dark matter results in a larger relic density for the same $\Treh$. For small ratios of $\Treh/M_*$ the resulting relic density scales as
\begin{equation} \label{eq:OmDMIEBH_scaling}
\Omega_{dm} \sim (0.08-0.3) \bigg(\frac{g_{dm} m_{dm}}{10\mathrm{~GeV}}\bigg)\bigg(\frac{M_{Pl}}{M_*}\bigg) \bigg(\frac{M_*}{20\,\Treh}\bigg)^{3/2} e^{20-M_*/\Treh} \,,
\end{equation}
where the range in the prefactor accounts for variation in $1 \le n \le 6$.

 The average momentum of dark matter at production is given by
\begin{equation}
    \pavg = \frac{\rho_{dm} (\Treh)}{n_{dm}(\Treh)}\,,
\end{equation}
where $\rho_{dm}$ is obtained via the same method as $n_{dm}$ except replacing $N_{dm}$ in \eq~\eqref{eq:ndmIEBHIntegral} with $E_{dm}$, the amount of energy produced by evaporation to dark matter from a single IEBH. In the relevant regime where $m_{dm} \ll \Tbh$,
\begin{align}
    E_{dm}(\Mbh) &= \int_{\Mbh}^0 d\Mbh^\prime \bigg(\frac{d\Mbh^\prime}{dt}\bigg)^{-1}\frac{dM_{\bullet\to dm}}{dt} \nonumber\\
    &= \frac{8\pi g_{dm} \tilde{\xi}_i}{(1+n)^2\alpha_0}\Mbh \,.
\end{align}
If after production, dark matter only loses energy due to the adiabatic expansion of the Universe, the average particle speed today would be
\begin{equation}
    \langle v_0 \rangle = \frac{g_{*s}(T_0)^{1/3} T_0 \pavg}{g_{*s}(\Treh)^{1/3} \Treh m_{dm}} \,.
\end{equation}
To leading order in $\Treh/M_*$, the average dark matter speed today can be approximated as
\begin{equation} \label{eq:vavgIEBH}
    \vavg \approx  \bigg(\frac{g_{*s}(T_0)^{1/3} M_* T_0 \tilde{\xi}_i}{g_{*s}(\Treh)^{1/3} m_{dm}\Treh \tilde{\eta}_i}\bigg)\bigg( \frac{(2+n)(8+5n)}{a_n(1+n)(7+5n)} \bigg).
\end{equation}
The average momentum of dark matter produced by IEBHs is independent of the dark matter mass. Therefore, heavier dark matter will be slower and colder today. Similarly, dark matter's temperature at formation is independent of the plasma temperature so for a constant scale of quantum gravity, a larger reheating temperature provides more time for dark matter to cool resulting in a lower average speed today.

\subsection{Planckeon dark matter}
Inspired by the black hole information paradox it was proposed that for black hole evaporation to maintain unitarity, the end product of evaporation must be quasi-stable objects called Planckeons\footnote{They are also sometimes referred to as ``Planck Relics'' or ``Black Hole Remnants''.} with $\Mbh\sim M_{Pl}$ and a lifetime much longer than the age of the Universe~\cite{Aharonov:1987tp}. This argument is supported by studies of black holes in quantum theories of gravity such as String Theory which show that quantum corrections to the Hawking temperature likely stops evaporation when black holes approach the Planck scale~\cite{Callan:1988hs,Alexeyev:2002tg}. Ref. \cite{Chen:2014jwq} provides a review of the many motivations for the existence of stable Planckeons as well as the theoretical arguments against their existence. In theories of LEDs, when the scale of quantum gravity is lower than the Planck scale, the arguments supporting stable Planckeons generalize so that they have a mass $\Mbh \sim M_*$ \cite{Hossenfelder:2003dy}.

The cosmological implications of stable Planckeons have been studied in four dimensions. With a reheating temperature close to the Planck scale, high-energy collisions can produce the correct relic abundance of Planckeons to account for all of the observed dark matter~\cite{Barrau:2019cuo,Friedlander:2023jzw}. Additionally, Planckeons generically have a non-zero charge, which has been shown to have important implications for direct detection prospects~\cite{Lehmann:2019zgt}. In this section we extend the analysis of Refs. \cite{Friedlander:2023jzw} and \cite{Barrau:2019cuo}, by generalizing it to theories of extra dimensions. 

The abundance of Planckeons is insensitive to the exact evolution of the black hole mass in the early Universe. Therefore, an analytic expression for the relic abundance of Planckeons can be found using the same method as described above for IEBHs. In the case of Planckeons, each formed black hole produces one Planckeon so that $N_{dm} =1$. This results in a number density of Planckeons, $n_{Pl}$, at formation of
\begin{equation}
    n_{Pl}(\Treh) = \frac{3\sqrt{\frac{5}{2}} a_n^2 g_*^{3/2}  M_{Pl} M_* (1+n)}{32(3+2n)\pi^4} \left[\left(\frac{\Treh}{M_*}\right)^\frac{5+3n}{1+n} \Gamma\left(\frac{13 + 9n}{2+2n}, \frac{M_*}{\Treh} \right) - \Gamma\left(\frac{1}{2}, \frac{M_*}{\Treh} \right)\right]\,. 
\end{equation}
Similarly, the density of Planckeons evolves in the same way as dark matter after formation.  So the relic abundance of Planckeons today, $\Omega_{Pl}$, can be obtained with \eq~\eqref{eq:OmDMIEBH} by assuming that all Planckeons have masses near the scale of quantum gravity, i.e. $m_{dm} = M_*$. Assuming that at the time of Planckeon formation, the plasma consists entirely of Standard Model particles, to leading order in $\Treh/M_*$, the relic density of Planckeons scales as
\begin{equation}
\Omega_{Pl} \sim (0.04-0.2) \bigg(\frac{M_*}{70\,\Treh}\bigg)^{3/2} e^{70-M_*/\Treh}\,,
\end{equation}
where the range in the prefactor accounts for all LED models with $0 \le n \le 6$ and the $70$ in the exponent denotes a benchmark ratio of $M_*/\Treh$ for which Planckeons comprise an $\mathcal{O}(1)$ fraction of the dark matter. One noteworthy feature of Planckeon production is that the dependence of the relic density today on $M_*$ and $\Treh$ is only in factors of $M_*/\Treh$. Therefore, the relic density today does not depend on the absolute scale of quantum gravity but only how large the reheating temperature is relative to it. 

\begin{figure}[htb]
    \centering
    \includegraphics[width=0.75\textwidth]{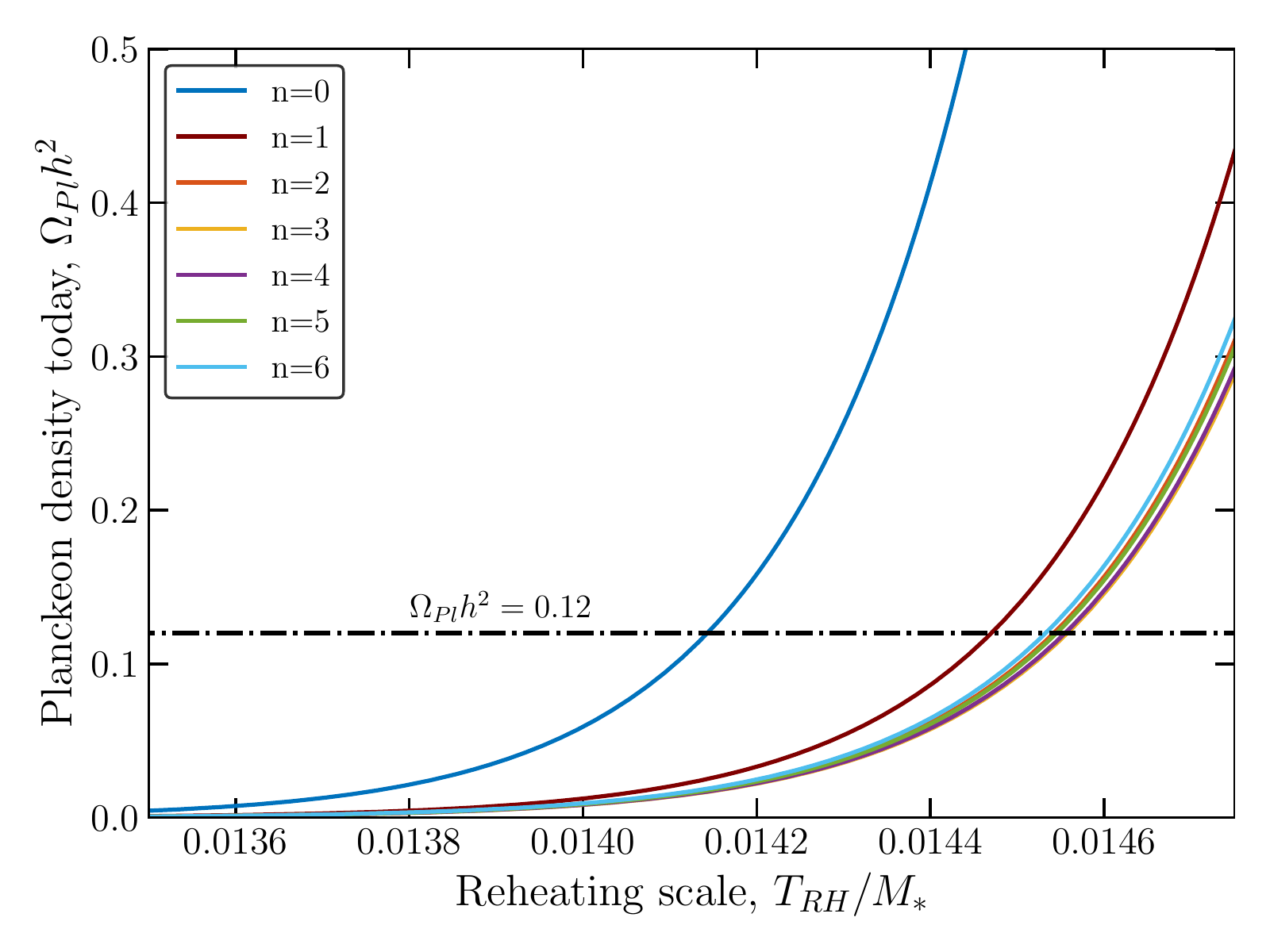}
    \caption{The relic abundance of Planckeons today if black holes with mass $\Mbh = M_*$ are stable. The black horizontal line shows the oberved relic abundance of dark matter while the different colour curves show the abundance of Planckeons with different numbers of LEDs. The relic abundance is independent of the scale of quantum gravity so this figure applies for all values of $M_*$.}
    \label{fig:planckRelicDensity}
\end{figure}

Figure \ref{fig:planckRelicDensity} shows the predicted density of Planckeons today as a function of the ratio $\Treh/M_*$ zoomed in on the parameter space where Planckeons comprise the entirety of dark matter. The blue $n=0$ curve shows the case of 4D Planckeons studied in Ref. \cite{Friedlander:2023jzw}. The abundance of Planckeons has a very strong dependence on $\Treh$ and is similar for all $n$. For a specific $M_\star$, changing the number of extra dimensions in the range of $0 \le n \le 6$ shifts the expected $\Treh$ value only by about $3\%$.

\section{Parameter Space and Observational Constraints} \label{sec:results}
In this section we connect observational constraints on dark matter to early Universe cosmology in the presence of LEDs. We focus on Dirac fermion dark matter so that the parameter space of our model is described entirely by four parameters: $n$, $M_*$, $\Treh/M_*$, and $m_{dm}$. Models of dark matter that have a different spin or number of degrees of freedom will have similar results as the case of Dirac fermion dark matter. Changes in the results stemming from different dark matter models can be accounted for through the values of $g_{dm}$, $\xi_i$, and $\eta_i$. We will present the results for the scenario of DEBHs and IEBHs separately, taking into account observational constraints, before directly comparing the two scenarios and demonstrating that IEBHs are dominant for the realistic production of dark matter.

\subsection{General considerations on dark matter constraints}
In determining the viability of black hole evaporation in the presence of LEDs as a dark matter production mechanism, we use three criteria: 1) the correct relic abundance of dark matter is produced, 2) the black holes evaporate before neutrino decoupling, and 3) the dark matter today is cold. For the relic abundance of dark matter, we use the {\it Planck} observation of $\Omega_{dm} h^2 = 0.12$~\cite{Planck:2018vyg}. While models which produce a smaller relic abundance are allowed, we are able to rule out parameter space which overproduces dark matter. For these excluded models, additional mechanisms would have to be introduced such as dark matter self-annihilation in order to wash out the produced abundance.

Limiting our scope to black holes which completely evaporate before neutrinos decouple from the plasma is not restrictive of IEBHs as they by definition immediately evaporate at $T\sim\Treh$. In principle, DEBHs could survive beyond neutrino decoupling and produce a cold dark matter candidate before the time of matter-radiation equality. However, LED black hole evaporation in this later epoch is severely constrained by its impact on Big Bang Nucleosynthesis (BBN)~\cite{Friedlander:2022ttk}. Therefore, our results will focus on black holes that evaporate before neutrino decoupling which do not contradict any cosmological observations. 

Measurements of the matter power spectrum via observations of the Lyman-$\alpha$ forest set constraints on the properties of warm dark matter (WDM). WDM washes out small scale structure, producing a deficit of power on those scales. An approximate lower bound on $m_{dm}$ can be found by ensuring that the average speed of dark matter is at most the maximum average speed of thermal WDM allowed by Lyman-$\alpha$ observations~\cite{Baldes:2020nuv}. The maximum average speed of thermal WDM today is~\cite{Bode:2000gq,Baldes:2020nuv}
\begin{equation} \label{eq:vdmMax}
    \vavg_{\rm max} \approx 3.9 \times 10^{-8} \bigg(\frac{1\,\mathrm{ keV}}{\mWDMmin} \bigg)^{4/3} \,,
\end{equation}
where $\mWDMmin = 4.65$~keV~\cite{Yeche:2017upn} is the minimum mass allowed for thermal WDM. We then set a lower bound on $m_{dm}$ by computing the average speed using \eq~\eqref{eq:pavgEvolution} in the case of DEBHs and \eq~\eqref{eq:vavgIEBH} in the case of IEBHs and ensuring $\vavg~\leq~\vavg_{\rm max}$. The impact of dark matter on the matter power spectrum depends on the full velocity-distribution of dark matter so this comparison to thermal WDM relies on an assumption that the non-thermal features of Hawking radiation do not have a large impact.

\subsection{Dark matter from Delayed Evaporating Black Holes}
\begin{figure}[htb]
    \centering
    \includegraphics[width=0.495\textwidth]{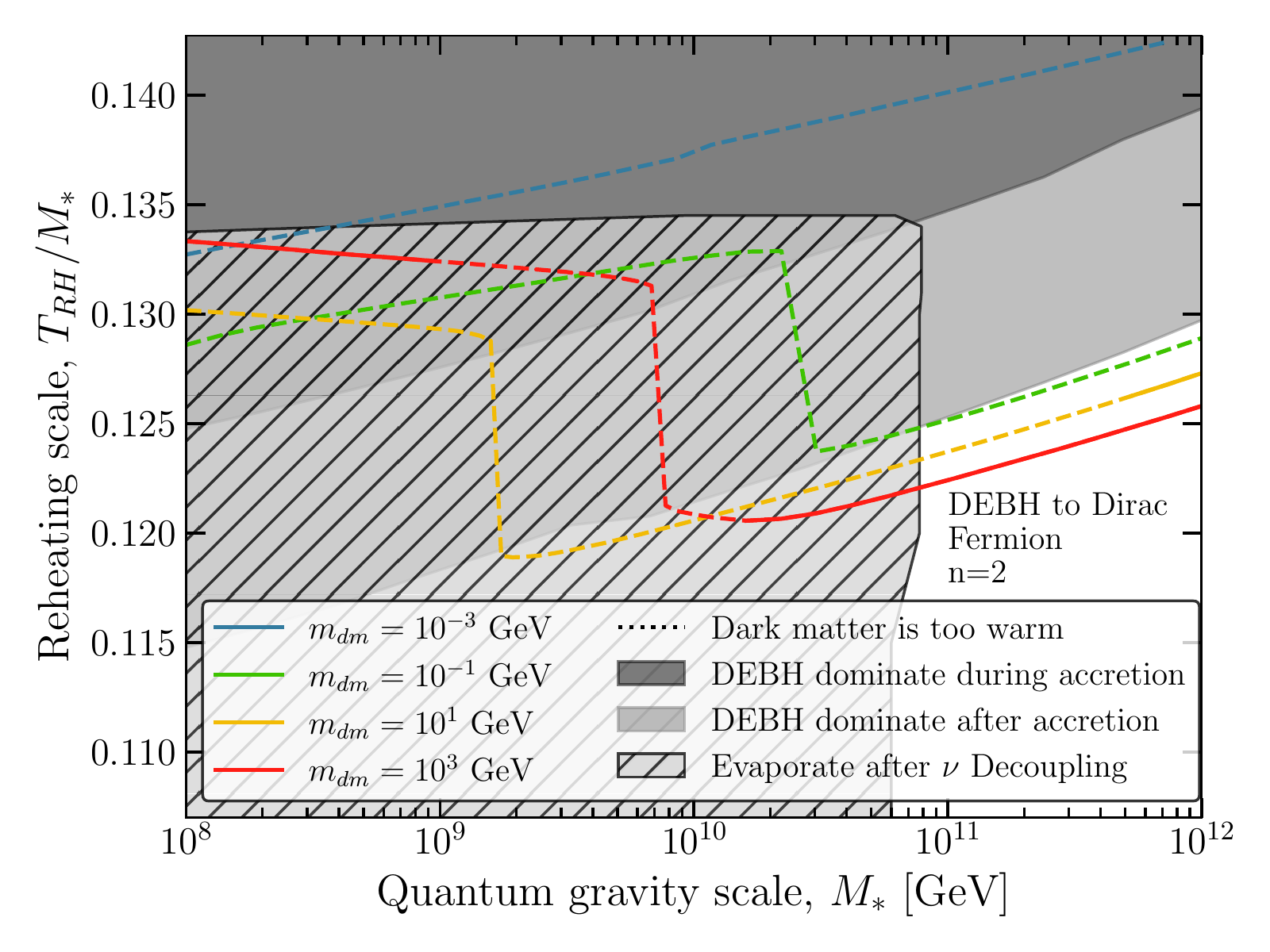}
    \includegraphics[width=0.495\textwidth]{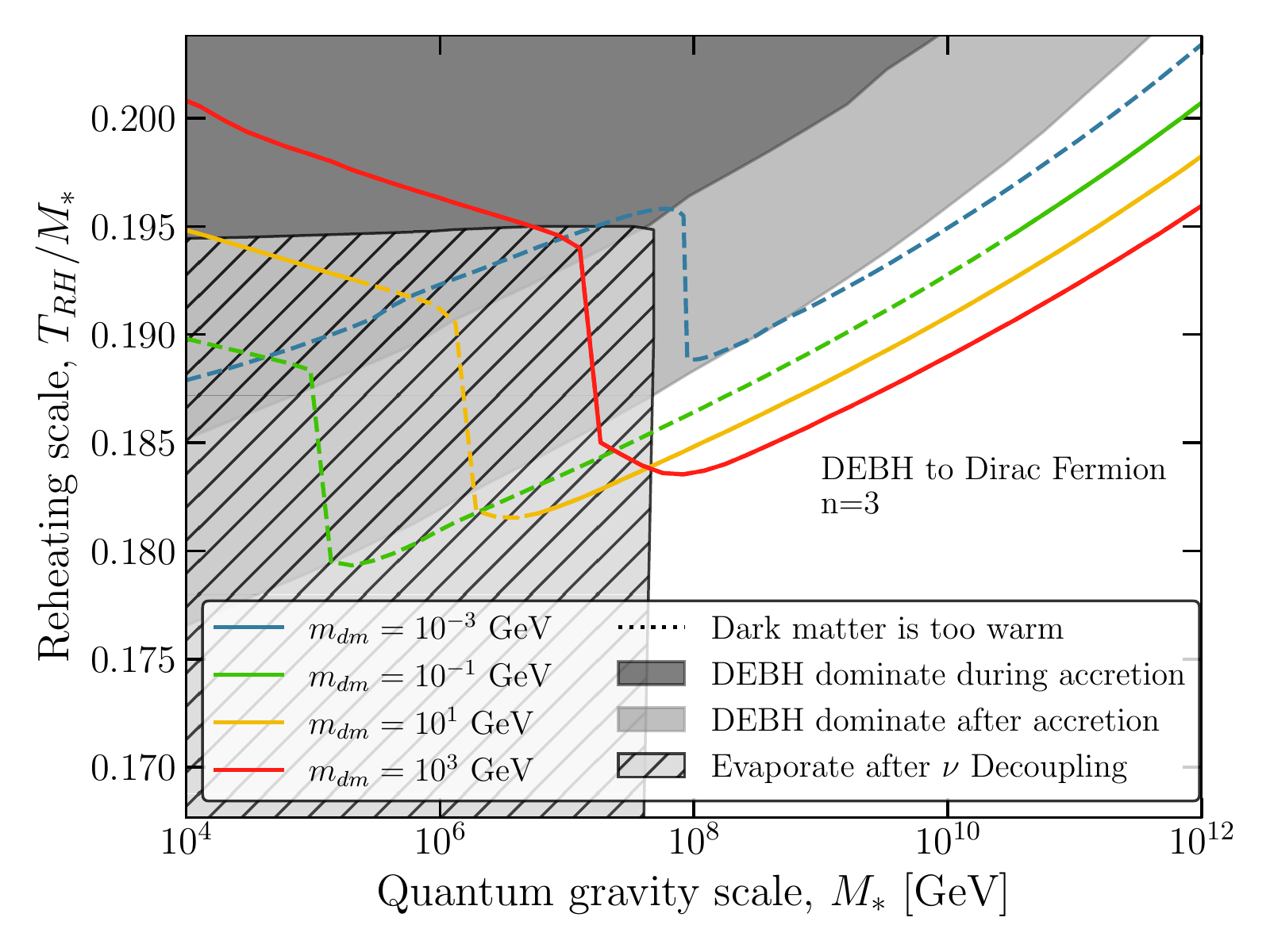}
    \includegraphics[width=0.495\textwidth]{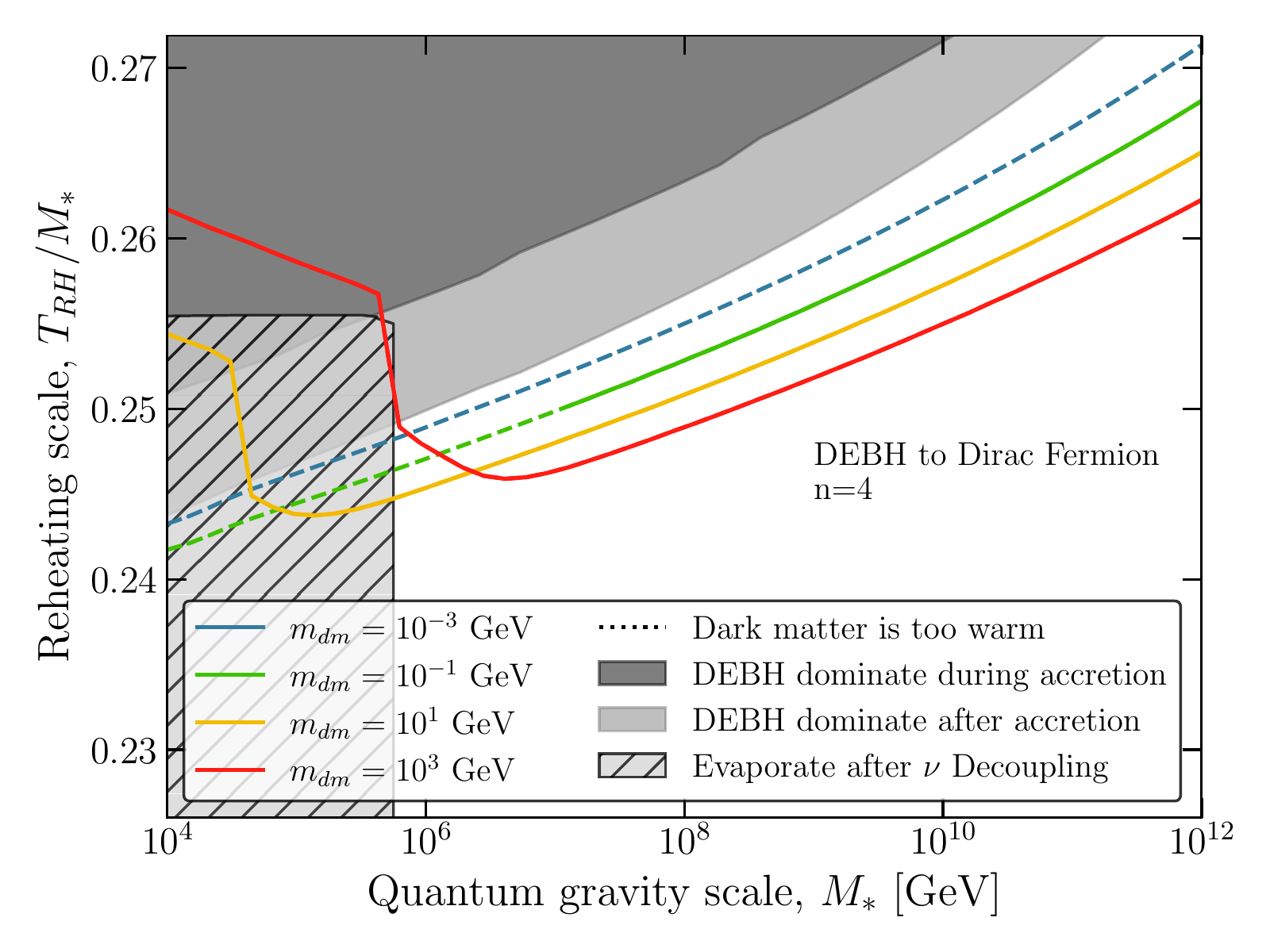}
    \includegraphics[width=0.495\textwidth]{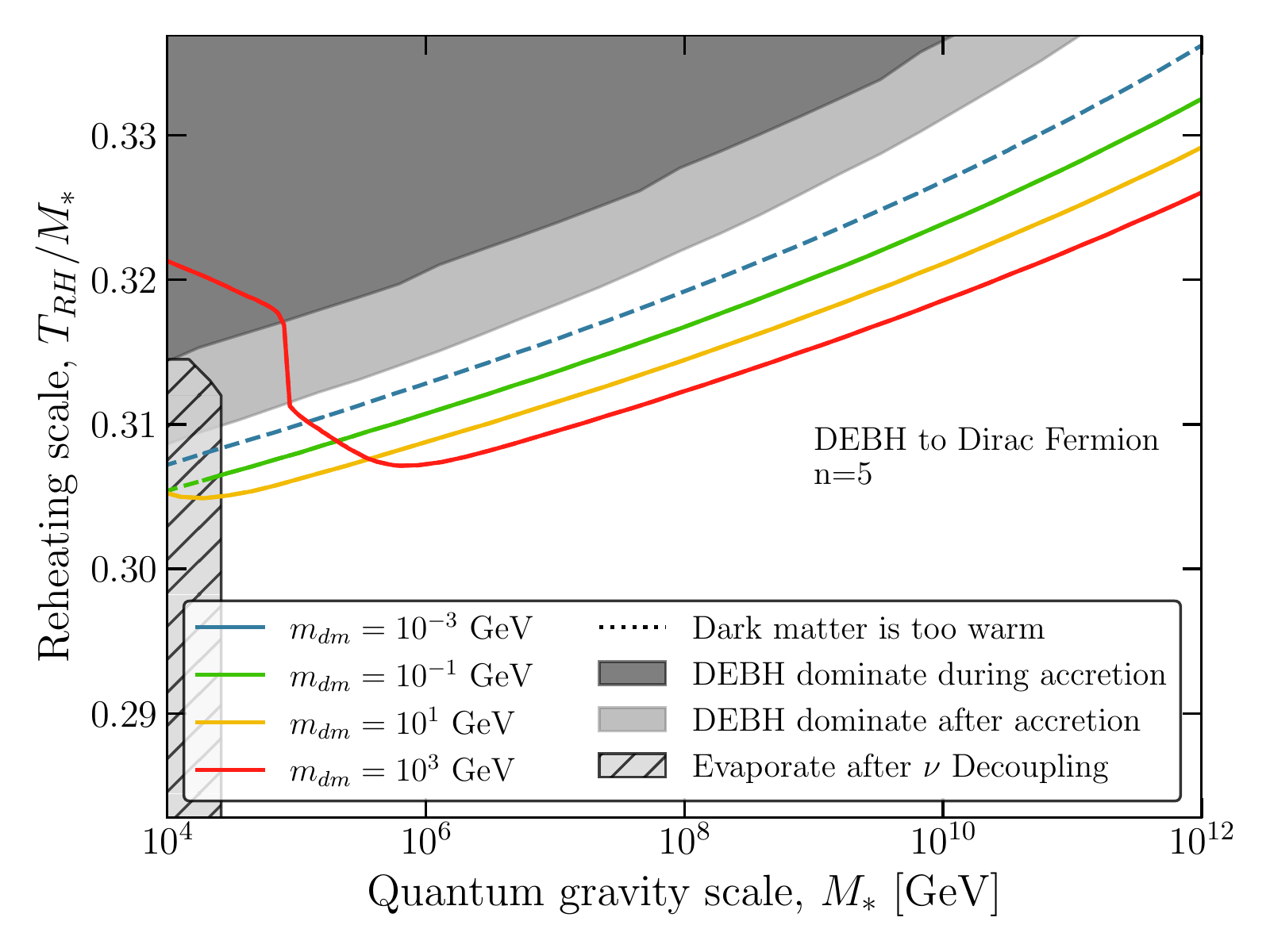}
    \includegraphics[width=0.495\textwidth]{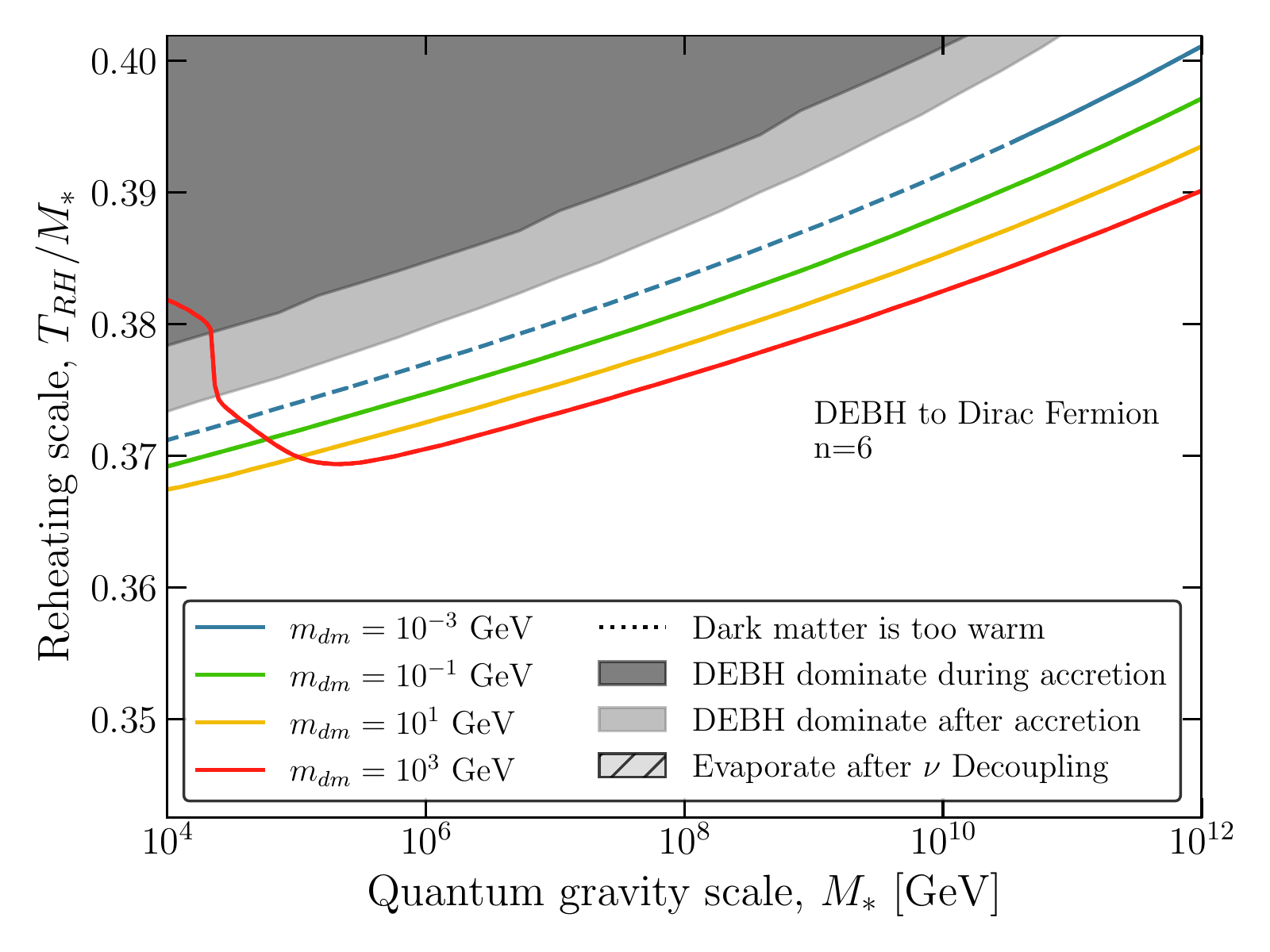}
    \caption{Reheating scale needed for DEBHs to produce the observed relic abundance of Dirac fermion dark matter as a function of the quantum gravity scale, $M_*$. Each panel depicts a different number of LEDs in the range $2 \le n \le 6$ and each curve depicts a different dark matter mass. The dashed portions of the curve show the parameter space where dark matter would be too warm in violation of observations of the Lyman-$\alpha$ forest. The grey hatched region shows parameter space where black holes survive until after neutrino decoupling. In the dark grey region, black hole growth is cutoff because DEBHs come to dominate the Universe before reaching their maximum mass. In the light grey region, DEBHs remain subdominant while growing but there exists a period of DEBH domination before they fully evaporate.}
    \label{fig:all_n_TRH_Mstar}
\end{figure}

The relationships required between the scale of quantum gravity and the reheating scale for DEBHs to produce the correct relic abundance of Dirac fermion dark matter are shown in Figure~\ref{fig:all_n_TRH_Mstar} for a variety of $m_{dm}$ and $n$ values. Each panel depicts a different number of LEDs and the different colour curves each depict a different value of $m_{dm}$. The shaded regions depict parameter space where there exists an era of early matter domination either during accretion (dark grey) or after accretion stops but before the DEBHs fully evaporate (light grey). The shaded regions are not to be interpreted as exclusion regions. The hatched region shows when DEBHs survive until after neutrino decoupling. In this region, black hole evaporation is constrained by the impact on BBN and the CMB~\cite{Friedlander:2022ttk}. However, previous work has not set constraints for black hole evaporation for all values of $M_*$ so DEBH evaporation to dark matter may remain viable for some portion of this parameter space.

The reheating scale for producing the observed relic abundance of dark matter from DEBHs is in the range $0.1 \le \Treh/M_* \le 0.4$ depending on the number of LEDs. If more LEDs exist, a larger $\Treh$ is required for the same scale of quantum gravity because DEBH production is heavily suppressed relative to models with fewer LEDs as seen in the left panel of Figure~\ref{fig:productionResults}. Models with fewer LEDs also produce larger DEBHs which have longer lifetimes. This means that for the produced DEBHs to entirely evaporate before neutrino decoupling, larger scales of quantum gravity are required. For example, in models with $n=2$ LEDs, for DEBHs to produce cold dark matter and entirely evaporation before neutrino decoupling, the LEDs must be small enough so that $M_* > 5\times 10^{10}$ GeV. Alternatively for $n=6$, all DEBHs with $M_* \geq 10^4$ GeV evaporate early, evading all BBN and CMB constraints.

The $\Treh/M_*$ ratios needed to produce the observed relic abundance of dark matter differ qualitatively for heavy and light dark matter. The trend for light dark matter can be seen for $n=3$ and $m_{dm} = 10^{-3}$ GeV. When $M_\star\gtrsim 10^8$~GeV, as $M_\star$ decreases the $\Treh/M_*$ ratio also decreases. This can be understood from the left panel of Figure \ref{fig:productionResults}, where a smaller $M_*$ requires a lower $\Treh/M_*$ ratio for the same fraction of the Universe to convert into DEBHs. There is a jump in the $\Treh/M_*$ ratio as $M_*$ drops below $M_*~\approx~10^8$~GeV with the curve entering the light grey shaded region. There, DEBHs grow to their maximum size during radiation domination but survive long enough so that they come to dominate the Universe before evaporating. In this regime, the black hole evaporation dumps a significant amount of energy into the plasma thus partially reheating it. This energy dump depletes the density of dark matter produced. Therefore, increases in $\Treh$ will cause more dark matter to be produced, but will also provide a larger radiation entropy dump which conversely dilutes the dark matter. This ultimately results in little change to the final relic abundance of dark matter. This trend stops in the dark grey region for large $\Treh$ values, when DEBHs are so abundant that they dominate the Universe during accretion. In this regime, the maximum mass that DEBHs grow to is cut off as accretion depletes the available radiation density. Dark matter is overproduced at small $M_\star$, and hence the $\Treh/M_*$ ratio decreases.

The behaviour for heavier dark matter models can be illustrated by the $m_{dm}=10^3$~GeV curve in the bottom ($n=6$) panel. Two differences can be spotted as compared with light dark matter, one being the uptick at $M_* \lesssim 10^5$~GeV before the jump and the other being the increase in $\Treh/M_*$ ratio in the dark grey region. The former is best explained by the right panel of Figure \ref{fig:productionResults}. So long as DEBHs do not dominate the Universe, models with smaller $M_*$ values produce larger and colder DEBHs. If $\Tbh \lesssim m_{dm}$, evaporation to dark matter is Boltzmann-suppressed. Therefore, these large DEBHs produced in models with a lower scale of quantum gravity produce very little dark matter, thus requiring more DEBHs to form and a larger $\Treh$.  This scenario changes in the dark grey region when the maximum black hole mass after accretion is reduced. A larger reheating temperature is also necessary to produce more and hence smaller and hotter black holes, so that dark matter production is not Boltzmann-suppressed, yielding the correct relic abundance today.

Generally, dark matter models with a lower mass require a larger $\Treh/M_*$ ratio to maintain the same relic abundance. This is because when $m_{dm} \ll \Tbh$ for DEBHs at their maximum mass, the energy-density of dark matter produced is independent of $m_{dm}$ but lighter dark matter is more energetic implying a smaller number density. Additionally, since the relic abundance today is proportional to $n_{dm} m_{dm}$, a lower dark matter mass requires an even larger number density for the same relic abundance. 

\subsection{Dark matter and Planckeons from Instantly Evaporating Black Holes}
\begin{figure}[t]
    \centering
    \includegraphics[width=0.49\textwidth]{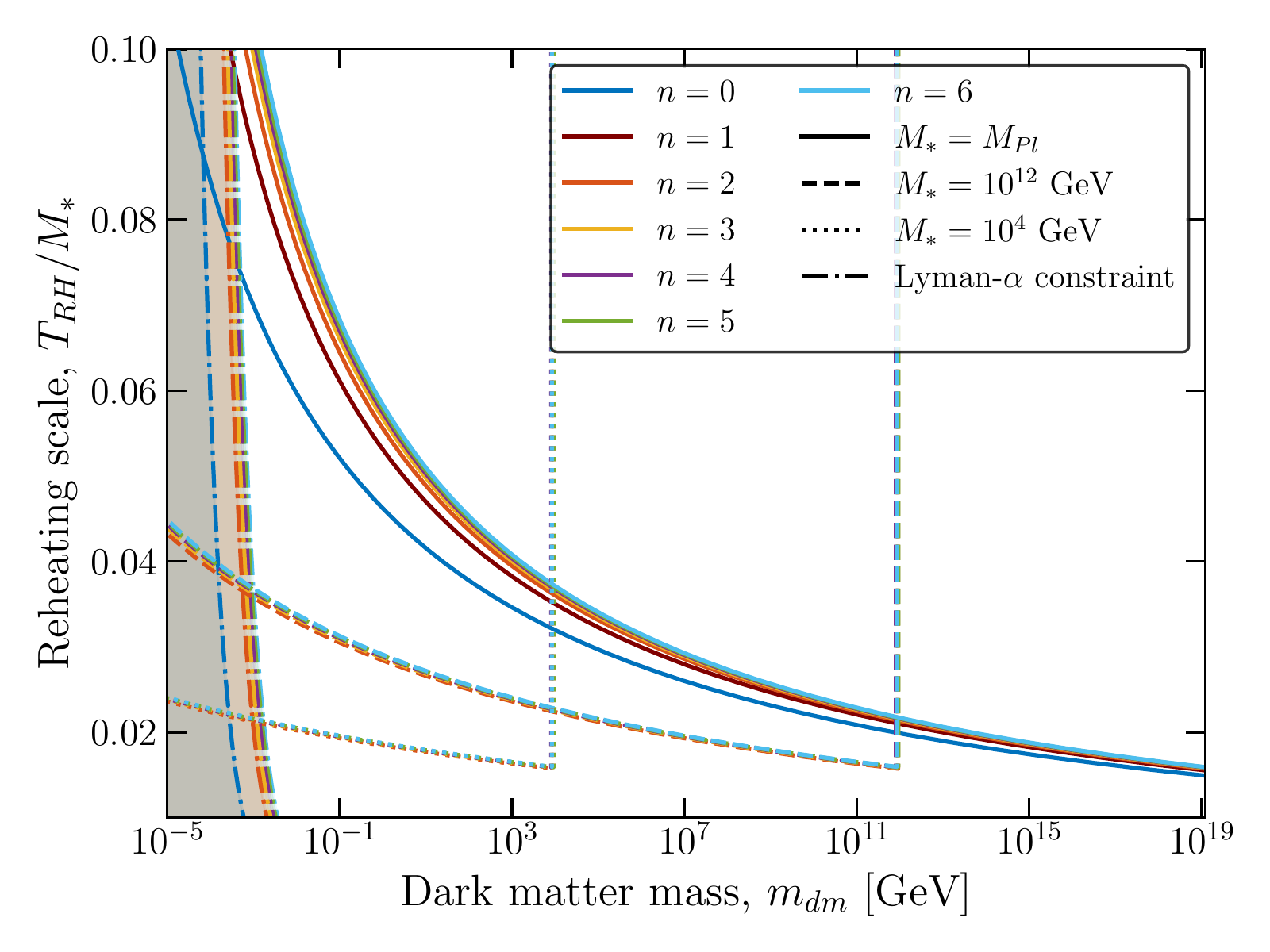}
    \includegraphics[width=0.49\textwidth]{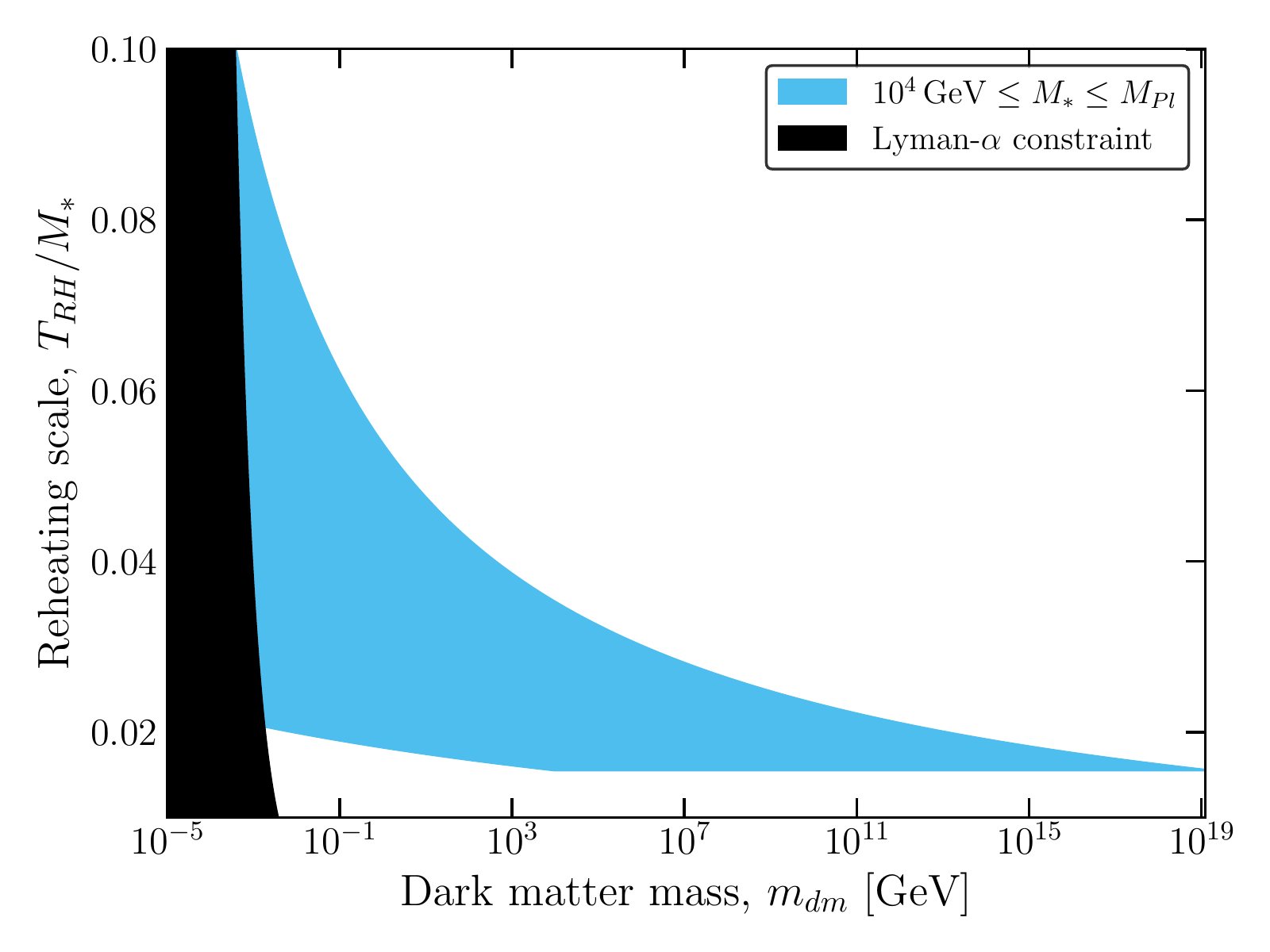}
    \caption{The reheating scale, $\Treh/M_*$, required for IEBHs to produce the observed relic abundance of dark matter as a function of dark matter mass, $m_{dm}$. Left: Each colour curve shows a different number of LEDs. The dark blue $n=0$ curve is the same as Ref.~\cite{Friedlander:2023jzw}. The solid, dashed, and dotted curves depict quantum gravity scales of $10^4$ GeV, $10^{12}$ GeV, and $M_*=M_{Pl}$ respectively. The dot-dashed curves show the minimum allowed dark matter mass, below which the produced dark matter would be in violation of Lyman-$\alpha$ forest constraints on warm dark matter. Right: The blue region shows the parameter space for which the entire relic abundance of dark matter can be produced by IEBHs with $n=6$ LEDs and $M_* \geq 10^4$ GeV. The black region is excluded by Lyman-$\alpha$ observations.}
    \label{fig:IEBHmdmTRH}
\end{figure}

Figure \ref{fig:IEBHmdmTRH} shows the reheating scale needed for IEBHs to produce the observed relic abundance of dark matter today as a function of $m_{dm}$ for various $M_*$ values. The different colour curves in the left panel represent different numbers of LEDs, demonstrating that varying $n$ in the range $1 \le n \le 6$ has very little impact on dark matter production. The solid, dashed, and dotted curves show $M_* = 10^4$ GeV, $M_* =10^{12}$ GeV, and $M_*=M_{Pl}$ respectively. The results in this figure use the analytic derivation of the relic abundance discussed in Section~\ref{sec:IEBHanyl}. This includes the assumption that $\Tbh \gg m_{dm}$ for most of the IEBHs which form. If instead, the abundance was obtained numerically without that assumption, the sharp cutoff at $M_*=m_{dm}$ would instead be a smooth cutoff due to the Boltzmann suppression of dark matter production. The dot-dashed curves represent the lowest allowed $m_{dm}$, below which dark matter would have an average speed that is too fast today and therefore be inconsistent with observations of the Lyman-$\alpha$ forest. By comparing the predicted average speed of dark matter produced by IEBHs Eq.~\eqref{eq:vavgIEBH} with the maximum average speed allowed for thermal WDM Eq.~\eqref{eq:vdmMax}, we find the constraint on the mass of fermionic dark matter produced by IEBHs to be
\begin{equation} \label{eq:IEBHmdmMin}
m_{dm} \gtrsim (1-3)\times 10^{-4}\,\mathrm{ GeV} \bigg(\frac{0.1 M_*}{\Treh}\bigg) \bigg(\frac{\mWDMmin}{4\,\mathrm{ keV}}\bigg)^{4/3} \,,
\end{equation}
where the range in the prefactor accounts for all LED models with $1 \le n \le 6$.

For $n\geq 2$, all scales of quantum gravity in the range $10^4\,\mathrm{GeV} \leq M_* \leq M_{Pl}$ are unconstrained. Therefore a region of the $m_{dm}-\Treh$ parameter space can be defined where the entire relic abundance of dark matter could be produced via IEBHs when varying $M_\star$. The right panel of Figure~\ref{fig:IEBHmdmTRH} shows this parameter space in blue. As the reheating temperature required to produce the correct relic abundance is very weakly dependent on $n$, we only show the region for $n=6$. The black region shows when the dark matter would be too light and warm, thus being excluded by observations of the Lyman-$\alpha$ forest. This figure demonstrates that in order to produce all of the dark matter, the ratio $\Treh/M_*$ must be between $10^{-2}$ and $10^{-1}$. For heavier dark matter models, the range of $\Treh$ values which are able to produce dark matter is smaller due to the implied constraint on quantum gravity such that $M_* > m_{dm}$.

\begin{figure}[htb]
    \centering
    \includegraphics[width=\textwidth]{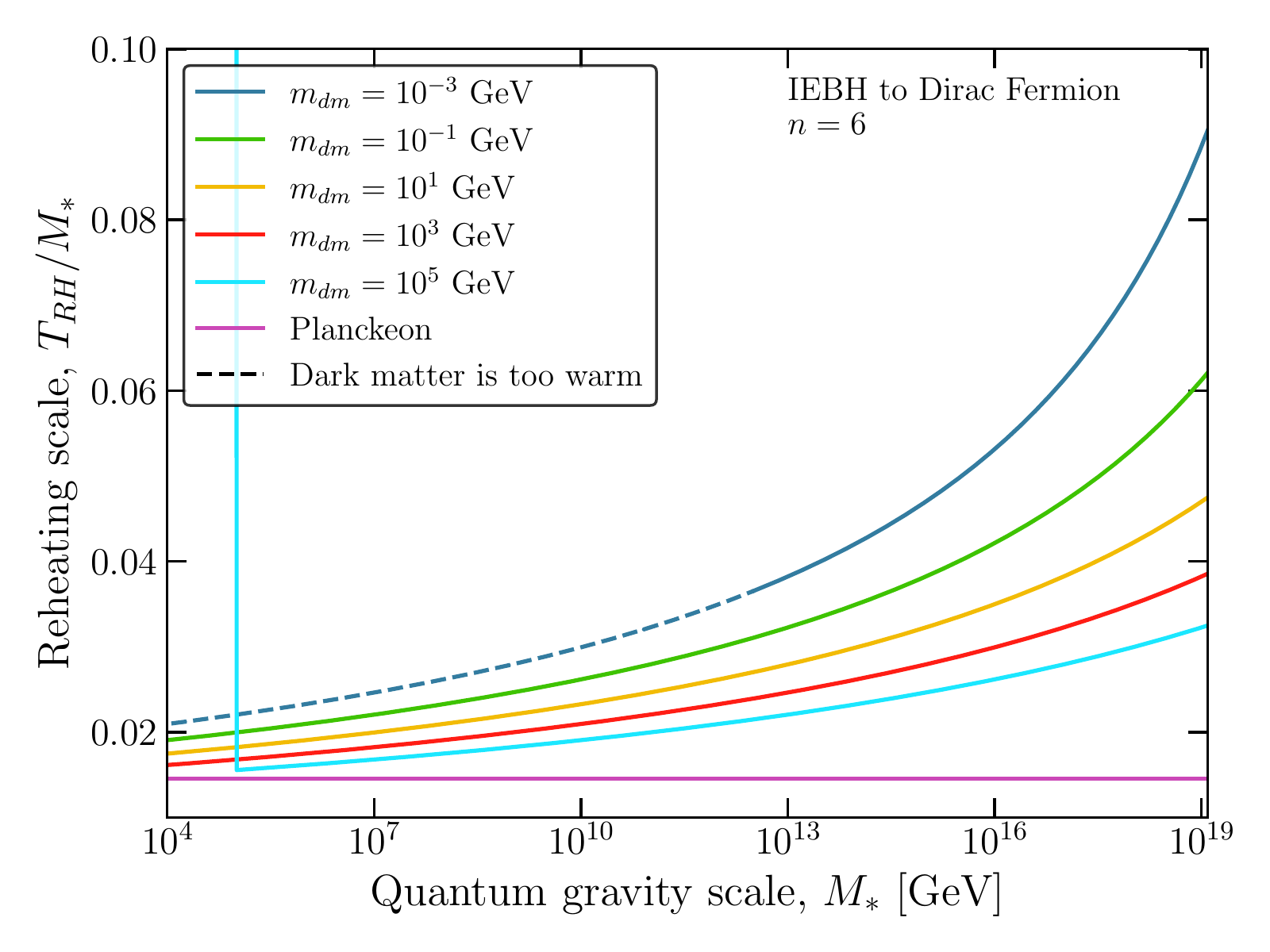}
    \caption{The reheating scale, $\Treh/M_*$, required to produce the observed relic abundance of dark matter as a function of the quantum gravity scale, $M_*$. The top five curves depict the scenario of IEBHs producing Dirac fermion dark matter with each curve showing a different dark matter mass. The dashed portion of the curves shows the parameter space where dark matter would be too warm today in violation of Lyman-$\alpha$ constraints on warm dark matter. The horizontal pink line shows the reheating scale required for Planckeons to comprise the entirety of dark matter today if black holes with mass $\Mbh=M_*$ are stable.}
    \label{fig:IEBHMstarTRH}
\end{figure}

Figure~\ref{fig:IEBHMstarTRH} shows the required reheating scale for IEBHs to produce the relic abundance of dark matter as a function of $M_*$ and compares this to the case of stable Planckeons. Only the case of $n=6$ is shown due to the weak dependence on $n$ for both IEBH production of dark matter and the production of stable Planckeons. Higher scales of quantum gravity require a $\Treh$ closer to $M_*$ because, as shown in \eq~\eqref{eq:OmDMIEBH_scaling}, the relic abundance of dark matter is enhanced by a factor of $M_{Pl}/M_*$. This enhancement arises from the fact that the Universe is assumed to obey the 4D Friedman equations so that the rate of the expansion of the Universe is suppressed by a factor of $M_{Pl}^{-1}$. For $M_{Pl} > M_*$, the Universe expands on a slower scale than IEBH production. This allows for more dark matter to be produced before the plasma cools sufficiently. The pink line shows that if black holes with mass $\Mbh = M_*$ are stable, the reheating scale required for Planckeons to comprise the entirety of dark matter is $\Treh = 1.5\times 10^{-2}\,M_*$. The other curves show different masses of Dirac fermion dark matter. The light blue curve, depicting $m_{dm} = 10^5$ GeV, is cutoff at $M_* = m_{dm}$. For all $m_{dm} \leq M_*$, the relic abundance of Planckeons is larger than that of Dirac fermion dark matter produced via IEBHs.  

\begin{figure}[htb]
    \centering
    \includegraphics[width=\textwidth]{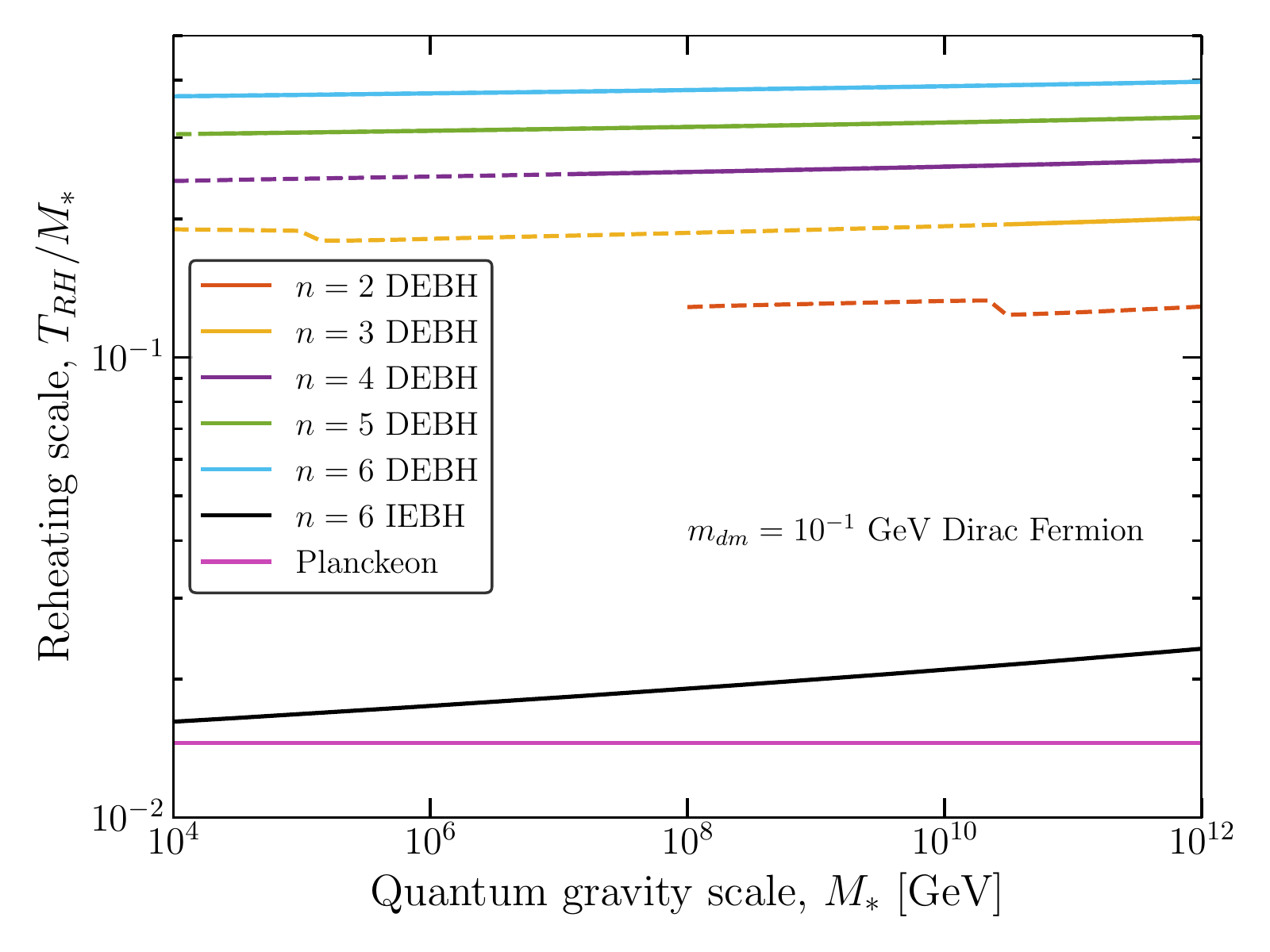}
    \caption{A comparison of the relevant reheating scale for each of the three production mechanisms discussed in this work: DEBH evaporation, IEBH evaporation, and Planckeon production. The top five curves show the reheating scale required for DEBHs to produce the observed relic abundance of Dirac fermion dark matter with a mass of $m_{dm} = 10^{-1}$ GeV for $2\leq n \leq 6$ LEDs. For the dashed segments of these curves, the produced dark matter would be too warm based on observations of the Lyman-$\alpha$ forest. The black curve shows the reheating scale for the scenario of IEBHs producing a $10^{-1}$ GeV mass Dirac fermion and the pink line shows the scenario where dark matter is comprised of stable Planckeons with mass $\Mbh=M_*$.}
    \label{fig:FullComparison}
\end{figure}

\subsection{Comparing IEBH and DEBH production}

By comparing Figures \ref{fig:all_n_TRH_Mstar} and \ref{fig:IEBHMstarTRH}, it is clear that the reheating scale required for DEBHs to produce the relic abundance of dark matter is higher than that for IEBHs. Figure~\ref{fig:FullComparison} makes this comparison explicit. The top five curves show the $\Treh/M_*$ scale required for DEBHs to produce the observed relic abundance of Dirac fermion dark matter with $m_{dm} = 10^{-1}$ GeV. Alternatively, the black line shows for the same dark matter model, the reheating scale required for IEBHs to produce the same relic abundance. Lastly, the horizontal pink line shows the case of stable Planckeons. 

Throughout this work we have treated IEBHs and DEBHs as two separate scenarios. In principle, both DEBHs and IEBHs can form and each produce a portion of the dark matter density. However, Figure~\ref{fig:FullComparison} demonstrates that for any cosmological model where $\Treh$ is large enough that DEBHs produce a significant portion of the dark matter, dark matter would be overabundant due to the much more efficient production via IEBHs. Conversely, at the temperatures where IEBHs do not overproduce dark matter, very few if any black holes form at masses large enough to be DEBHs. This is true for all scales of quantum gravity and dark matter models we have studied. Therefore, it is unlikely that DEBH evaporation is the source of the observed dark matter abundance. 

Our calculation of the dark matter production via IEBHs relies on a semiclassical treatment of the formation and evaporation of $M_*$-scale black holes. These results are therefore sensitive to quantum gravitational effects. DEBHs on the other hand typically form and produce dark matter in a regime where the semiclassical treatment is trustworthy. Therefore, studying DEBHs provides an upper limit on the required reheating scale for dark matter produciton via LED black holes which is robust to UV effects in quantum gravity.  

\section{Conclusions} \label{sec:conclusions}
In this work we have demonstrated that in the presence of LEDs and a reheating temperature within a factor of $10^{-2}-10^{-1}$ of the reduced scale of quantum gravity, energetic collisions in the early Universe would produce microscopic black holes that evaporate, leaving behind a relic abundance of dark matter. This scenario does not rely on any non-gravitational interactions between dark matter and the Standard Model. We have compared three mechanisms for dark matter production via black holes: 1) DEBHs where black holes grow via accretion before ultimatley evaporating, 2) IEBHs where $M_*$-scale black holes immediately evaporate, and 3) Planckeons where dark matter is comprised of stable black hole relics with $\Mbh = M_*$. For any $\Treh$ where DEBHs are relevant, dark matter would be vastly overproduced by IEBHs or as Planckeons. Therefore, IEBHs and Planckeons are the two viable mechanisms for producing the observed relic abundance of dark matter depending on whether $M_*$ mass black holes are stable or not. The production of dark matter via IEBHs and Planckeons is very similar to previous work which studied these mechanisms in a standard 4D cosmology~\cite{Barrau:2019cuo,Friedlander:2023jzw}. However, in a 4D cosmology a reheating temperature of $\Treh \gtrsim 10^{17}$ GeV is required which is not allowed in single-field inflation models~\cite{Planck:2018jri}. In this work the requirement of a hot big bang is removed by postulating the existence of additional compactified spatial dimensions. 

Strong yet model dependent constraints have been set on $\Treh$ in LED models based on the production of Kaluza Klein (KK) modes~\cite{Hannestad:2001nq}. These constraints would exclude all $\Treh$ where black holes are produced via energetic collisions. However, these constraints were determined assuming that the LEDs are compactified to a toroidal geometry such that there exist light and long-lived KK modes. If the LEDs have a different geometry, such as a compact hyperbolic manifold, these cosmological bounds are entirely evaded. Furthermore, the presence of additional branes \cite{Hannestad:2001nq} or couplings with matter \cite{Illana:2020jpi} would cause the KK modes to have a short lifetime, removing the cosmological constraints. Another caveat that has recently been pointed out regarding black hole formation in the presence of LEDs is that if the Compton wavelength is altered similarly to the black hole horizon radius, the true scale of quantum gravity may remain at the Planck scale~\cite{Lake:2018hyv,Lake:2023nng}.

As we consider LED models with scales of quantum gravity far out of the reach of colliders and with compactification length scales much smaller than can be tested in the lab, it is important to consider potential observational signatures that can discover this production mechanism. As shown in Table~\ref{tab:xiEtaValues}, evaporation to gravitons is significantly enhanced in the presence of LEDs. This enhancement is caused by gravitons being able to travel in the bulk. Therefore, the majority of this radiation will be in the form of KK modes rather than massless gravitational waves. The impact of evaporation to KK modes is dependent on their mass, lifetime, and couplings which in turn depend on the compactification scheme of the LEDs. Studying the impact of the evaporation to KK modes and gravitational waves may provide unique observational signals. An intriguing possibility is that long-lived KK modes could be a cold dark matter candidate, removing the need to postulate the existence of an additional species (see Refs. \cite{Cheng:2002ej,Servant:2002aq} in the context of a different extra dimension model). 
Even if evaporation to KK modes is suppressed by the LED compactification, evaporation to gravitational waves on the brane may produce a detectable signal of gravitational microwaves. The energy density of gravitational waves is depleted quicker than that of dark matter making observations of gravitational waves from dark matter evaporation sensitive to smaller $M_*$ values. We leave a study of the observational signatures of IEBHs in specific LED compactification models to future work. 

We have demonstrated a novel impact of LED models independent of the specific geometry. In doing so, we has expanded the scope of connections between primordial black holes and dark matter and have illustrated a rich phenomenology of LED models that has not yet been explored.

\section*{Acknowledgements}
AF is supported by an Ontario Graduate Scholarship. NS is supported by the National Natural Science Foundation of China (NSFC) Project No. 12047503. NS also  acknowledges the UK Science and Technology Facilities Council for support through the Quantum Sensors for the Hidden Sector collaboration under the grant ST/T006145/1. ACV is supported by the Arthur B.~McDonald Canadian Astroparticle Physics Research Institute, NSERC, and the province of Ontario via an Early Researcher Award. Equipment is funded by the Canada Foundation for Innovation and the Province of Ontario, and housed at the Queen's Centre for Advanced Computing. Research at Perimeter Institute is supported by the Government of Canada through the Department of Innovation, Science, and Economic Development, and by the Province of Ontario.

\appendix

\section{Dark Matter Momentum Evolution Derivation} \label{app:momEvolDeriv}
Using $\rho_{mom}\equiv n \pavg$. In an isolated collisionless system:
\begin{align}
a^{-4}\frac{d(a^4 \rho_{mom})}{dt} &= a^{-4}\bigg(a\pavg \frac{d(a^3 n)}{dt} + a^3 n \frac{d(a\pavg)}{dt}\bigg)\\
&= 0
\end{align}
Now introducing a ``collision term" which accounts for black hole evaporation to dark matter with the assumption that the dark matter mass is a negligible contribution to its total energy at the time of evaporation
\begin{align}
a^{-4}\frac{d(a^4 \rho_{mom})}{dt} &= C_{\bullet\to dm}\\
&=n_\bullet \bigg|\frac{dM_{\bullet\to dm}}{dt}\bigg|\\
\frac{d(\rho_{mom})}{dt} &= n_\bullet \bigg|\frac{dM_{\bullet\to dm}}{dt}\bigg| - 4a^{-1} \rho_{mom}\dot{a}\\
\frac{d(n\pavg)}{dt} &= - 4H n \pavg + n_\bullet \bigg|\frac{dM_{\bullet\to dm}}{dt}\bigg| 
\end{align}

\bibliography{LEDBH.bib}
\end{document}